\documentclass[aps,prb,twocolumn,twoside,preprintnumbers,amsmath,amssymb,superscriptaddress,floatfix]{revtex4}

\usepackage[utf8]{inputenc}
\usepackage{mathptm}  
\usepackage{dcolumn}                    
\usepackage{bm}                        
\usepackage{graphicx}
\usepackage{times}
\usepackage{epstopdf}
\usepackage{color}
\usepackage{subfig}
\usepackage{soul}

\usepackage{tikz}
\usepackage{float}

\newcommand{\ie}{\textit{i.e. }}	
\newcommand{\eg}{\textit{e.g. }}	
\newcommand{\etal}{\emph{et al.}}

\newcommand{\pauli}[1]{\sigma_{#1}}
\renewcommand{\v}[1]{\underline{#1}}
\newcommand{\uv}[1]{\underline{\hat{#1}}}
\renewcommand{\d}{\mathrm{d}}

\newcommand{\p}[1]{{#1}^{\vphantom{\dagger}}}
\newcommand{\cc}[1]{{#1}^*}

\newcommand{\up}{\uparrow}
\newcommand{\dn}{\downarrow}
\newcommand{\pll}{{\mkern3mu\vphantom{\perp}\vrule depth 0pt\mkern2mu\vrule depth 0pt\mkern3mu}}

\newcommand{\re}[1]{\mathrm{Re}\{#1\}}

\newcommand{\abs}[1]{\vert#1\vert}
\newcommand{\expect}[1]{\langle#1\rangle}
\newcommand{\comm}[2]{[#1,\,#2]}
\newcommand{\anticomm}[2]{\{#1,\,#2\}}

\newcommand{\figref}[1]{Fig.~\ref{fig:#1}}
\newcommand{\figrefs}[2]{Figs.~\ref{fig:#1}~and~\ref{fig:#2}}

\newcommand{\eqnref}[1]{Eq.~\eqref{#1}}
\newcommand{\eqnrefs}[2]{Eqs.~\eqref{#1}~and~\eqref{#2}}
\newcommand{\refcite}[1]{Ref.~\onlinecite{#1}}
\newcommand{\appref}[1]{Appendix~\ref{app:#1}}
\newcommand{\secref}[1]{Section~\ref{sec:#1}}

\newcommand{\cd}{\tilde{\nabla}}
\newcommand{\rn}{N}
\newcommand{\rnn}{\tilde{N}}
\newcommand{\rg}{\gamma}
\newcommand{\rgx}{(\partial_z\rg)}
\newcommand{\rgxx}{(\partial_z^2\rg)}
\newcommand{\rgg}{\tilde{\gamma}}
\newcommand{\rggx}{(\partial_z\rgg)}
\newcommand{\rggxx}{(\partial_z^2\rgg)}
\newcommand{\rrn}[1]{N_{#1}}
\newcommand{\rrnn}[1]{\tilde{N_{#1}}}
\newcommand{\rrg}[1]{{\gamma}_{#1}}
\newcommand{\rrgx}[1]{(\partial_z\rrg{#1})}

\newcommand{\rrgg}[1]{\tilde{\gamma}_{#1}}
\newcommand{\rrggx}[1]{(\partial_z\rrgg{#1})}

\renewcommand{\i}{i}

\begin{document}

\title{Critical Temperature and Tunneling Spectroscopy of Superconductor/Ferromagnet \\
Hybrids with Intrinsic Rashba--Dresselhaus Spin-Orbit Coupling}

\author{Sol H. Jacobsen,$^{\!1*}$ Jabir Ali Ouassou,$^{\!1*}$ and Jacob Linder}
\affiliation{Department of Physics, Norwegian University of
Science and Technology, N-7491 Trondheim, Norway \\
${}^{*}$These authors contributed equally to this work.}


\begin{abstract}
\noindent
We investigate theoretically how the proximity effect in superconductor/ferromagnet hybrid structures with intrinsic spin-orbit coupling manifests in two measurable quantities, namely the density of states and critical temperature.
To describe a general scenario, we allow for both Rashba and Dresselhaus type spin-orbit coupling. 
Our results are obtained via the quasiclassical theory of superconductivity, extended to include spin-orbit coupling in the Usadel equation and Kupriyanov--Lukichev boundary conditions.
Unlike previous works, we have derived a Riccati parametrization of the Usadel equation with spin-orbit coupling which allows us to address the full proximity regime and not only the linearized weak proximity regime.
First, we consider the density of states in both SF~bilayers and SFS~trilayers, where the spectroscopic features in the latter case are sensitive to the phase difference between the two superconductors.
We find that the presence of spin-orbit coupling leaves clear spectroscopic fingerprints in the density of states due to its role in creating spin-triplet Cooper pairs.
Unlike SF and SFS structures without spin-orbit coupling, the density of states in the present case depends strongly on the direction of magnetization. Moreover, we show that the spin-orbit coupling can stabilize spin-singlet superconductivity even in the presence of a strong exchange field $h \gg \Delta$. This leads to the possibility of a magnetically tunable minigap: changing the direction of the exchange field opens and closes the minigap.
We also determine how the critical temperature $T_c$ of an SF bilayer is affected by spin-orbit coupling and, interestingly, demonstrate that one can achieve a spin-valve effect with a single ferromagnet.
We find that $T_c$ displays highly non-monotonic behavior both as a function of the magnetization direction and the type and direction of the spin-orbit coupling, offering a new way to exert control over the superconductivity of proximity structures.

\end{abstract}

\maketitle


\section{Introduction}
Material interfaces in hybrid structures give rise to proximity effects, whereby the properties of one material can ``leak'' into the adjacent material, creating a region with properties derived from both materials.
In superconductor/ferromagnet (SF) hybrid structures \cite{buzdin_rmp_05}, the proximity effect causes superconducting correlations to penetrate into the ferromagnetic region and vice versa.
These correlations typically decay over short distances, which in diffusive systems is of the order $\sqrt{D/h}$, where $D$ is the diffusion coefficient of the ferromagnet and $h$ is the strength of the exchange field.
However, for certain field configurations, the singlet correlations from the superconductor may be converted into so-called long-range triplets (LRTs) \cite{bergeret_rmp_05}.
These triplet components have spin projection parallel to the exchange field, and decay over much longer distances.
This results in physical quantities like supercurrents decaying over the length scale $\xi_N = \sqrt{D/T}$, which is usually much larger than the ferromagnetic coherence length $\xi_F = \sqrt{D/h}$, where $T$ is the temperature.
This distance is independent of $h$, and at low temperatures it becomes increasingly large, which allows the condensate to penetrate deep into the ferromagnet.
The isolation and enhancement of this feature has attracted much attention in recent years as it gives rise to novel physics and possible low-temperature applications by merging spintronics and superconductivity \cite{linder_nphys_15}.

It is by now well-known that the conversion from singlet to long-range triplet components of the superconducting state can happen in the presence of magnetic inhomogeneities \cite{bergeret_prl_01, eschrig_prl_03}, \ie a spatially varying exchange field, and until recently such inhomogeneities were believed to be the primary source of this conversion \cite{halterman, cottet, golubov, millo, robinson, birge, alidoust,shomali_njp_12, trifunovic_prb_10, sosnin_prl_06}, although other proposals using \eg non-equilibrium distribution functions and intrinsic triplet superconductors also exist \cite{annunziata_prb_11, bobkova,konschelle_prb_10, houzet_prl_07}. However, it has recently been established that another possible source of LRT correlations is the presence of a finite spin-orbit (SO) coupling, either in the superconducting region \cite{annunziata_prb_12} or on the ferromagnetic side \cite{BergeretTokatly2013,BergeretTokatly2014}.
In fact, it can be shown that an SF structure where the magnetic inhomogeneity is due to a Bloch domain wall, as considered in \eg Refs.~\onlinecite{eschrig_adv, burmistrov, linder_prb_14}, is gauge equivalent to one where the ferromagnet has a homogeneous exchange field and intrinsic SO coupling\cite{BergeretTokatly2014}.
It is known that SO scattering can be caused by impurities\cite{AbrikosovGorkov1962}, but this type of scattering results in purely isotropic spin-relaxation, and so does not permit the desired singlet-LRT conversion.
To achieve such a conversion, one needs a rotation of the spin pair into the direction of the exchange field \cite{Eschrig2011}.
This can be achieved by using materials with an intrinsic SO coupling, either due to the crystal structure in the case of noncentrosymmetric materials\cite{Samokhin2009}, or due to interfaces in thin-film hybrids\cite{Edelstein2003}, where the latter also modifies the fundamental process of Andreev reflection \cite{LinderYokoyama2011, sun_prb_15}. The role of SO coupling with respect to the supercurrent in ballistic hybrid structures has also been studied recently \cite{helix}.

In this paper, we establish how the presence of spin-orbit coupling in SF structures manifests in two important experimental observables: the density of states~$D(\epsilon)$ probed via tunneling spectroscopy (or conductance measurements), and the critical temperature~$T_c$.
A common consequence for both of these quantities is that neither becomes independent of the magnetization direction.
This is in contrast to the case without SO coupling in conventional monodomain ferromagnets, where the results are invariant with respect to rotations of the magnetic exchange field.
This symmetry is now lifted due to SO coupling: depending on the magnetization direction, LRT Cooper pairs are created in the system which leave clear fingerprints both spectroscopically and in terms of the $T_c$ behavior. 
On the technical side, we will present in this work for the first time a Riccati parametrization of the Usadel equation and its corresponding boundary conditions that include SO coupling.
This is an important advance in terms of exploring the full physics of triplet pairing due to SO coupling as it allows for a solution of the quasiclassical equations without any assumption of a weak proximity effect. We will also demonstrate that the SO coupling can actually protect the singlet superconducting correlations even in the presence of a strong exchange field, leading to the possibility of a minigap that is magnetically tunable via the orientation of the exchange field.

The remainder of the article will be organised as follows: In Section \ref{Sec:Theory}, we introduce the relevant theory and notation, starting from the quasiclassical Usadel equation, which describes the diffusion of the superconducting condensate into the ferromagnet. We also motivate our choice of intrinsic SO coupling in this section, and propose a new notation for describing Rashba--Dresselhaus couplings.
The section goes on to discuss key analytic features of the equations in the limit of weak proximity, symmetries of the density of states at zero energy, and analytical results needed to calculate the critical temperature of hybrid systems.
We then present detailed numerical results in Section~\ref{Sec:Results}:
we analyze the density of states of an SF bilayer in \ref{SubSec:SF} [see Fig.~\ref{Fig:model}(a)], with the case of pure Rashba coupling considered in Section~\ref{SubSec:PureR}, and we study the SFS Josephson junction in \ref{SubSec:SFS} [see Fig.~\ref{Fig:model}(b)].
We consider different orientations and strengths of the exchange field and SO coupling, and in the case of the Josephson junction, the effect of altering the phase difference between the condensates.
Then, in Section~\ref{sec:results-Tc}, we continue our treatment of the SF~bilayer in the full proximity regime by including a self-consistent solution in the superconducting layer, and focus on how the presence of  SO coupling affects the critical temperature of the system.
We discover that the SO coupling allows for spin-valve functionality with a single ferromagnetic layer, meaning that rotating the magnetic field by $\pi/2$ induces a large change in $T_c$.
Finally, we conclude in Section~\ref{Sec:Disc} with a summary of the main results, a discussion of some additional consequences of the choices made in-text, as well as possibilities for further work.

\section{Theory}\label{Sec:Theory}
\subsection{Fundamental concepts}
The diffusion of the superconducting condensate into the ferromagnet can be described by the Usadel equation, which is a second-order partial differential equation for the Green's function of the system\cite{Usadel1970}.
Together with appropriate boundary conditions, the Usadel equation establishes a system of coupled differential equations that can be solved in one dimension.
We will consider the case of diffusive equilibrium, where the retarded component $\hat{g}^R$ of the Green's function is sufficient to describe the behaviour of the system \cite{belzig_review_98, chandrasekhar_review_03}.
We start by examining the super\-conducting correlations in the ferromagnet, and use the standard Bardeen--Cooper--Schrieffer (BCS) bulk solution for the superconductors.
In particular, we will clarify the spectroscopic consequences of having SO coupling in the ferromagnetic layer. 

\begin{figure}[H]
  \centering
  \includegraphics[width=0.8\linewidth]{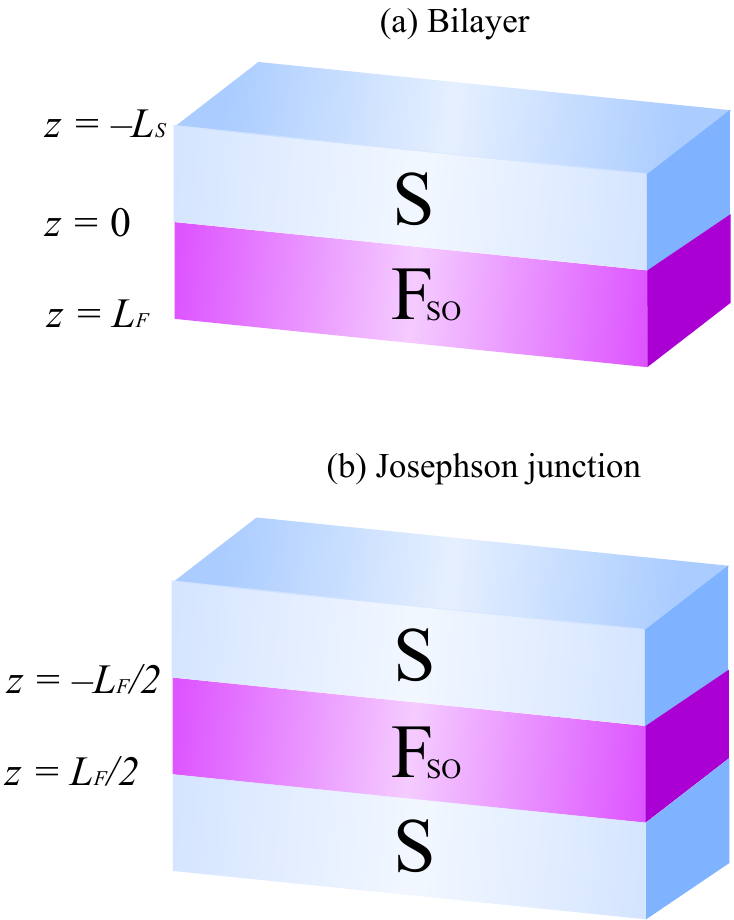}
  \caption
   {(Color online) (a) The SF bilayer in \ref{SubSec:SF}, \ref{SubSec:PureR} and \ref{sec:results-Tc}. We take the thin-film layering direction along the $z$-axis, and assume an $xy$-plane Rashba--Dresselhaus coupling in the ferromagnetic layer. (b) The SFS trilayer in \ref{SubSec:SFS}.}
	\label{Fig:model}
\end{figure}

In the absence of SO coupling, the Usadel equation \cite{Usadel1970} in the ferromagnet reads
\begin{eqnarray}\label{Eqn:Usadel1}
	D_F\nabla(\hat{g}^R\nabla \hat{g}^R)+i\left[\epsilon\hat{\rho}_3+\hat{M}, \hat{g}^R\right]=0,
\end{eqnarray}
where the matrix $\hat{\rho}_3=\textrm{diag}(1,-1)$, and $\epsilon$ is the quasiparticle energy.
The magnetization matrix $\hat M$ in the above equation is
\begin{eqnarray}
	\hat{M}=
	\begin{pmatrix}
	\underline{h}\cdot\underline{\sigma} & 0\\
	0 & (\underline{h}\cdot\underline{\sigma})^*
	\end{pmatrix},\nonumber
\end{eqnarray}
where $\underline{h}=(h_x,h_y,h_z)$ is the ferromagnetic exchange field, $(^*)$ denotes complex conjugation, $\underline{\sigma} = (\pauli x, \pauli y, \pauli z)$ is the Pauli vector, and $\pauli k$ are the usual Pauli matrices.
The corresponding Kupriyanov--Lukichev boundary conditions are\cite{KuprianovLukichev1988}
\begin{eqnarray}\label{Eqn:KL1}
	2L_j\zeta_j\hat{g}^R_j\nabla\hat{g}^R_j=[\hat{g}^R_1,\hat{g}^R_2] \, ,
\end{eqnarray}
where the subscripts refer to the different regions of the hybrid structure; in the case of an SF bilayer as depicted in Fig.~\ref{Fig:model}(a), $j=1$ denotes the superconductor, and $j=2$ the ferromagnet, while $\nabla$ denotes the derivative along the junction $1\to2$.
The respective lengths of the materials are denoted $L_j$, and the interface parameters $\zeta_j = R_B/R_j$ describe the ratio of the barrier resistance $R_B$ to the bulk resistance $R_j$ of each material.

We will use the Riccati parameterisation\cite{SchopohlMaki1995} for the quasi\-classical Green's function~$\hat{g}^R$,
\begin{eqnarray}\label{Eqn:gR}
	\hat{g}^R=
	\begin{pmatrix}
	N(1+\gamma\tilde{\gamma}) & 2N\gamma\\
	-2\tilde{N}\tilde{\gamma} & -\tilde{N}(1+\tilde{\gamma}\gamma)
	\end{pmatrix},
\end{eqnarray}
where the normalisation matrices are $N=(1-\gamma\tilde{\gamma})^{-1}$ and $\tilde{N}=(1-\tilde{\gamma}\gamma)^{-1}$.
The tilde operation denotes a combination of complex conjugation $i\rightarrow -i$ and energy $\epsilon\rightarrow -\epsilon$, with $\gamma\rightarrow \tilde{\gamma}$, $N\rightarrow \tilde{N}$. The Riccati parameterisation is particularly useful for numerical computation because the parameters are bounded $[0,1]$, contrary to the multi-valued $\theta$-parameterisation \cite{belzig_review_98}. In practice, this means that for certain parameter choices the numerical routines will only converge in the Riccati formulation. Appendix~\ref{app:riccati} contains some further details on this parameterisation.

To include intrinsic SO coupling in the Usadel equation, we simply have to replace all the derivatives in \eqnref{Eqn:Usadel1} with their gauge covariant counterparts:\cite{gorini_prb_10, BergeretTokatly2014}
\begin{eqnarray}
	\nabla(\,\cdot\,) \mapsto \tilde{\nabla}(\,\cdot\,) \equiv \nabla(\,\cdot\,) -i[\v{\hat A},\;\cdot\;] \, .
\end{eqnarray}
This is valid for any SO coupling linear in momentum.
We consider the leading contribution; higher order terms, \eg those responsible for the SU(2) Lorentz force, are neglected here. Such higher order terms are required to produce so-called $\varphi_0$ junctions which have attracted interest of late\cite{phi0}, and consequently we will see no signature of the $\varphi_0$ effect in the systems considered herein. The object $\v{\hat A}$ has both a vector structure in geometric space, and a $4\times4$ matrix structure in Spin--Nambu space, and can be written as $\v{\hat A} = \mathrm{diag}(\v A,-\v A^*)$ in terms of the SO field $\v A = (A_x,A_y,A_z)$, which will be discussed in more detail in the next subsection. SO coupling in the context of quasiclassical theory has also been discussed in Refs.~\onlinecite{konschelle_epb_14, gorini_prb_10}.
When we include the SO coupling as shown above, we derive the following form for the Usadel equation (see \appref{riccati}):
\begin{eqnarray}\label{Eqn:SOUsadel}
	\lefteqn{D_F\left(\partial_k^2 \gamma + 2(\partial_k \gamma)\tilde{N}\tilde{\gamma}(\partial_k \gamma)\right)} \nonumber\\
	&=& -2i\epsilon\gamma - i\underline{h} \cdot (\underline{\sigma}\gamma-\gamma\underline{\sigma}^*)\nonumber\\
	&&\,+D_F\left[\underline{A}\underline{A}\gamma-\gamma\underline{A}^*\underline{A}^*+2(\underline{A}\gamma+\gamma\underline{A}^*)\tilde{N}(\underline{A}^*+\tilde{\gamma}\underline{A}\gamma)\right]\nonumber\\
	&&\,+2iD_F\left[(\partial_k \gamma)\tilde{N}({A}^*_k+\tilde{\gamma}{A}_k\gamma)+({A}_k+\gamma {A}^*_k\tilde{\gamma})N(\partial_k \gamma)\right]\!,
\end{eqnarray}
where the index $k$ indicates an arbitrary choice of direction in Cartesian coordinates.
The corresponding equation for $\tilde{\gamma}$ is found by taking the tilde conjugate of Eq.~(\ref{Eqn:SOUsadel}).
Similarly, the boundary conditions in \eqnref{Eqn:KL1} become:
\begin{eqnarray}
	\partial_k\gamma_1=\frac{1}{L_1\zeta_1}(1-\gamma_1\tilde{\gamma}_2)N_2(\gamma_2-\gamma_1)+i{A}_k\gamma_1+i\gamma_1{A}_k^*,\nonumber\\
	\partial_k\gamma_2=\frac{1}{L_2\zeta_2}(1-\gamma_2\tilde{\gamma}_1)N_1(\gamma_2-\gamma_1)+i{A}_k\gamma_2+i\gamma_2{A}_k^*, \label{Eqn:KLRic}
\end{eqnarray}
and the $\tilde{\gamma}$ counterparts are found in the same way as before. For the details of these derivations, see \appref{riccati}.

We will now discuss the definition of current in the presence of spin-orbit interactions. Since the Hamiltonian including SO coupling contains terms linear in momentum (see below), the velocity operator $v_j = \partial H/\partial k_j$ is affected. We stated above that the Kupriyanov-Lukichev boundary conditions are simply modified by replacing the derivative with its gauge covariant counterpart including the SO interaction. To make sure that current conservation is still satisfied, we must carefully examine the Usadel equation. In the absence of SO coupling, the quasiclassical expression for electric current is given by
\begin{align}
	\v{\,I\!}_{\,e} = I_0 \int^\infty_{-\infty} \text{d}\varepsilon\, \text{Tr}\{ \rho_3(\check{g}\nabla\check{g})^K \},
\end{align}
where $\check{g}$ is the $8\times8$ Green's function matrix in Keldysh space
\begin{align}
\check{g} = \begin{pmatrix}
\hat{g}^R & \hat{g}^K \\
\hat{0} & \hat{g}^A \\
\end{pmatrix} ,
\end{align}
and $I_0$ is a constant that is not important for this discussion. Current conservation can now be proven from the Usadel equation itself. We show this for the case of equilibrium, which is relevant for the case of supercurrents in Josephson junctions. In this case $\hat{g}^K = (\hat{g}^R - \hat{g}^A)\tanh(\varepsilon/2T)$ and we get
\begin{align}
\v{\,I\!}_{\,e} = I_0 \int^\infty_{-\infty} \text{d}\varepsilon\, \text{Tr}\{ \rho_3(\hat{g}^R\nabla\hat{g}^R - \hat{g}^A\nabla\hat{g}^A)\}\tanh(\varepsilon/2T).
\end{align}
Performing the operation Tr$\{\rho_3 \cdots\}$ on the Usadel equation, we obtain
\begin{align}
D \nabla \cdot \mathrm{Tr}\{\rho_3 (\hat{g}^R \nabla \hat{g}^R ) + \i\, \mathrm{Tr}\{\rho_3[\varepsilon\rho_3 + \hat{M},\hat{g}^R]\} = 0.
\end{align}                              
Now, inserting the most general definition of the Green's function $\hat{g}^R$, one finds that the second term in the above equation is always zero. Thus, we are left with
\begin{align}
\nabla \cdot \mathrm{Tr}\{\rho_3 (\hat{g}^R \nabla \hat{g}^R )\} = 0 \,,
\end{align}
which expresses precisely current conservation since the same analysis can be done for $\hat{g}^A$. Now, let us include the SO coupling. The current should then be given by
\begin{align}
\v{\,I\!}_{\,e} = I_0 \int^\infty_{-\infty} \text{d}\varepsilon\, \text{Tr}\{ \rho_3(\check{g}\tilde{\nabla}\check{g})^K \} \,,
\end{align}
so that the expression for the charge current is modified by the presence of SO coupling, as is known. The question is now if this current is conserved, as it has to be physically. We can prove that it is from the Usadel equation by rewriting it as
\begin{align}
D\nabla \cdot &(\hat{g}^R \tilde{\nabla}\hat{g}^R) \notag\\
 &= D[\v{A}, \hat{g}^R\nabla\hat{g}^R] + D[\v{A},[\v{A},\hat{g}^R]] - \i[\varepsilon\rho_3 + \hat{M},\hat{g}^R] \,,
\end{align}
and then performing the operation Tr$\{\rho_3 \cdots\}$, one finds:
\begin{align}
D\nabla \cdot \text{Tr}\{\rho_3(\hat{g}^R \tilde{\nabla}\hat{g}^R)\} = 0,
\end{align}
so we recover the standard current conservation law $\nabla \cdot \v{\,I\!}_{\,e}=0$.

\subsection{Spin-orbit field}\label{SubSec:Theory_SO}
The precise form of the generic SO field $\v{A}$ is imposed by the experimental requirements and limitations.
As the name suggests, spin-orbit coupling couples a particle's spin with its motion, and more specifically its momentum.
As mentioned in the Introduction, the SO coupling in solids can originate from a lack of inversion symmetry in the crystal structure. Such spin-orbit coupling can be of both Rashba and Dresselhaus type and is determined by the point group symmetry of the crystal \cite{dresselhaus, socbook}. It is also known that the lack of inversion symmetry due to surfaces, either in the form of interfaces to other materials or to vacuum, will give rise to antisymmetric spin-orbit coupling of the Rashba type. For sufficiently thin structures, the SO coupling generated in this way can permeate the entire structure, but the question of precisely how far into adjacent materials such surface-SO coupling may penetrate appears to be an open question in general.
Intrinsic inversion asymmetry arises naturally due to interfaces between materials in thin-film hybrid structures such as the ones considered herein.
Noncentrosymmetric crystalline structures provide an alternative source for intrinsic asymmetry, and are considered in \refcite{JacobsenLinder2015}.
In thin-film hybrids, the Rashba spin splitting derives from the cross product of the Pauli vector $\underline{\sigma}$ with the momentum $\underline{k}$,
\begin{equation}
	\p H_{R} = -\frac{\alpha}{m}(\underline{\sigma}\times \underline{k})\cdot \hat{\underline{z}} \, ,
\end{equation}
where $\alpha$ is called the Rashba coefficient, and we have chosen a coordinate system with $\hat{\underline{z}}$ as the layering direction.
Another well-known type of SO coupling is the Dresselhaus spin splitting, which can occur when the crystal structure lacks an inversion centre.
For a two-dimensional electron gas (quantum well) confined in the $\hat{\underline z}$-direction, then to first order $\langle k_z \rangle = 0$, so the Dresselhaus splitting becomes
\begin{equation}\label{eq:dhaus}
	\p H_{D} = \frac{\beta}{m}(\sigma_y k_y - \sigma_x k_x) \, ,
\end{equation}
where $\beta$ is called the Dresselhaus coefficient. In our structure, we consider a thin-film geometry with the confinement being strongest in the $z$-direction. Although there may certainly be other terms contributing to the Dresselhaus SO coupling in such a structure, since real thin-film structures will have three-dimensional quasiparticle diffusion and we use a $2D$ form of the SO coupling here, we consider the standard form Eq. (\ref{eq:dhaus}) as an approximation that captures the main physics in the problem.
This is a commonly used model in the literature to explore the effects originating from SO coupling in a system.
When we combine both interactions, we obtain the Hamiltonian for a general Rashba--Dresselhaus SO coupling,
\begin{equation}
	\p H_{RD} = \frac{k_x}{m}(\alpha\pauli y - \beta\pauli x ) - \frac{k_y}{m}(\alpha\pauli x - \beta\pauli y) \, .
\end{equation}
In this work, we will restrict ourselves to this form of SO coupling. It should be noted that our setup may also be viewed as a simplified model for a scenario where the SO coupling and ferromagnetism exist in separate, thin layers, in which case we expect qualitatively similar results to the ones reported in this manuscript. 

As explained in \refcite{BergeretTokatly2014}, the SO coupling acts as a background SU(2) field, \ie an object with both a vector structure in geometric space, and a $2\times2$ matrix structure in spin space.
We can therefore identify the interaction above with an effective vector potential $\v A$ which we will call the \emph{SO field},
\begin{equation}
	\p H_{RD}\equiv - \underline{k}\cdot\underline{A}/m \, ,
\end{equation}
from which we derive that
\begin{eqnarray}
	\underline{A}=(\beta \sigma_x-\alpha\sigma_y, \alpha\sigma_x-\beta\sigma_y,0) \, .\label{Eqn:A}
\end{eqnarray}

\begin{figure}[H]
	\hspace{0.5em}
	\begin{tikzpicture}
		\draw[<->,thick]  (2.5,2.0) node[above] {\hspace{1.7em}$\pauli x \cos \chi + \pauli y \sin \chi$}
                               -- (0.0,0.0)
                               -- (3.2,0.0) node[right] {$k_x$};
    		\draw[blue,thick] (1.0,0.0) arc (0:20:1.0) node[right] {$\chi$} [->] arc (20:35:1.0);
		\draw[<->,thick]  (4.5,+2.5) node[above] {$k_y$}
                               -- (4.5,-0.7) 
                               -- (6.5,+1.8) node[above] {\hspace{1.5em}$\pauli x \sin \chi + \pauli y \cos \chi$};
    		\draw[blue,thick] (4.5,0.3) arc (90:70:1.0) node[above] {$\chi$} [->] arc (70:55:1.0);
	\end{tikzpicture}
	\caption{Geometric interpretation of  the SO field (\ref{Eqn:SOpolar}) in polar coordinates: the Hamiltonian couples the momentum component $k_x$ to the spin component $(\pauli x \cos \chi + \pauli y \sin \chi)$ with a coefficient $+a/m$, and the momentum component $k_y$ to the spin component ($\pauli x \sin \chi + \pauli y \cos \chi$) with a coefficient $-a/m$. Thus, $a$ determines the magnitude of the coupling, and $\chi$ the angle between the coupled momentum and spin components.}
	\label{fig:soc-geometry}
\end{figure}
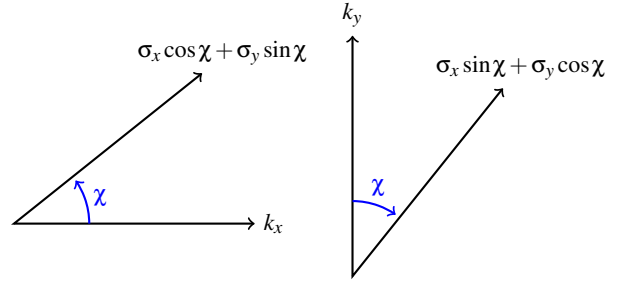

At this point, it is convenient to introduce a new notation for describing Rashba--Dresselhaus couplings, which will let us distinguish between the physical effects that derive from the strength of the coupling, and those that derive from the geometry.
For this purpose, we employ polar notation defined by the relations
\begin{align}
	\alpha &\equiv          -   a \sin \chi \, , \nonumber\\
	\beta  &\equiv \phantom{-}\!a \cos \chi \, , 
\end{align}
where we will refer to $a$ as the \emph{SO strength}, and $\chi$ as the \emph{SO angle}.
Rewritten in the polar notation, \eqnref{Eqn:A} takes the form:
\begin{equation}\label{Eqn:SOpolar}
	\v A = a(\pauli x \cos \chi + \pauli y \sin \chi) \uv x - a(\pauli x \sin \chi + \pauli y \cos \chi) \uv y \; .
\end{equation} 
From the definition, we can immediately conclude that $\chi = 0$ corresponds to a pure Dresselhaus coupling, while $\chi = \pm\pi/2$ results in a pure Rashba coupling, with the geometric interpretation of $\chi$ illustrated in \figref{soc-geometry}.
Note that $A_x^2 = A_y^2 = a^2$, which means that $\v A^2 = 2a^2$.
Another useful property is that we can switch the components $A_x \leftrightarrow A_y$ by letting $\chi \rightarrow 3\pi/2-\chi$.

The appearance of LRTs in the system depends on the interplay between SO coupling and the direction of the exchange field. Recall that the LRT components are defined as having spin projections parallel to the exchange field, as opposed to the short-ranged triplet (SRT) component which appears as long as there is exchange splitting\cite{FFLO} but has spin projection perpendicular to the field and is therefore subject to the same pair-breaking effect as the singlets\cite{Eschrig2011,linder_nphys_15}, penetrating only a very short distance into strong ferromagnets.
Thus if we have an SO field component along the layering direction, \eg if we had ${A}_z\neq 0$ in Figs.~\ref{Fig:model}(a) and \ref{Fig:model}(b), achievable with a noncentrosymmetric crystal or in a nanowire setup, then a non-vanishing commutator $\comm{\underline{A}}{\v h \cdot \v \sigma}$ creates the LRT.
However, we will from now only consider systems where ${A}_z=0$, in which case the criterion for LRT is \cite{BergeretTokatly2014} that $\comm{\underline{A}}{\comm{\underline{A}}{\v h \cdot \v \sigma}}$ must not be parallel to the exchange field $\v h \cdot \v \sigma$. Expanding, we have
\begin{align}
	\comm{\v A}{\comm{\v A}{\v h\cdot\v \sigma}} 	&=   4a^2 (\,\v h \cdot \v \sigma\, + h_z\pauli z) 	\nonumber\\
							&\;- 4a^2 (h_x\pauli y + h_y \pauli x ) \sin 2\chi 	\; ,
	\label{Eqn:cond}
\end{align}
from which it is clear that no LRTs can be generated for a pure Dresselhaus coupling $\chi = 0$ or Rashba coupling $\chi = \pm \pi/2$ when the exchange field is in-plane.
However, the effect of SO coupling becomes increasingly significant for angles close to $\pm\pi/4$ (see Fig.~\ref{Fig:DoSAngle} in Section~\ref{SubSec:SF}).
We also see that no LRTs can be generated for in-plane magnetization in the special case $h_x = h_y$ and $h_z = 0$, since $h_x\pauli y + h_y \pauli x$ can then be rewritten as $h_x \pauli x + h_y \pauli y$, which is parallel to $\v h$.
There is no LRT generation for the case $h_x = h_y = 0$ and $h_z \neq 0$ for similar reasons.
In general however, the LRT will appear for an in-plane magnetization as long as $h_x\neq h_y$ and the SO coupling is not of pure Dresselhaus or pure Rashba type. It is also important to note that the LRT can be created even for pure Rashba type SO coupling if the magnetization has both in- and out-of-plane magnetization components. We will discuss precisely this situation in Sec. \ref{SubSec:PureR}.

Once the condition for long-range triplet generation is satisfied, increasing the corresponding exchange field will also increase the proportion of long-range triplets compared with short-range triplets.
Whether or not the presence of long-range triplets can be observed in the system, \ie if they retain a clear signature in measurable quantities such as the density of states when the criteria for their existence is fulfilled, depends on other aspects such as the strength of the spin-orbit coupling and will be discussed later in this paper. \textit{Thus, a main motivation for this work is to take a step further than discussing their existence \cite{BergeretTokatly2014} and instead make predictions for when long-ranged triplet Cooper pairs can actually be observed via spectroscopic or $T_c$ measurements in SF structures with spin-orbit coupling.} However, we will also demonstrate that the presence of SO coupling offers additional opportunities besides the creation of LRT Cooper pairs. We will show both analytically and numerically that the SO coupling can protect the singlet component even in the presence of an exchange field, which normally would suppress it. This provides the possibility of tuning the well-known minigap \textit{magnetically}, both in bilayer and Josephson junctions, simply by altering the direction of the magnetization.

\subsection{Weak proximity effect}\label{SubSec:WPE}
In order to establish a better analytical understanding of the role played by SO coupling in the system before presenting the spectroscopy and $T_c$ results, we will now consider the limit of weak proximity effect, which means that $|\gamma_{ij}|\ll 1$, $N\approx 1$ in the ferromagnet.
The anomalous Green's function in general is given by the upper-right block of Eq. (\ref{Eqn:gR}), $f = 2N\gamma$, which we see reduces to $f = 2\gamma$ in this limit.
It will also prove prudent to express the anomalous Green's function using a singlet/triplet decomposition, where the singlet component is described by a scalar function $f_s$, and the triplet components encapsulated in the so-called $d$-vector\cite{BalianWerthamer1963, MackenzieMaeno2003},
\begin{equation}
	f = (f_s + \v d \cdot \v \sigma) i\pauli y \, .
\end{equation}
Combining the above with the weak proximity identity $f = 2\gamma$, we see that the components of $\gamma$ can be rewritten as:
\begin{equation}
	\label{Eqn:wpegamma}
      	\gamma = \frac12 \begin{pmatrix} id_y - d_x &  d_z + f_s  \vspace{0.4em}\\
                       			d_z - f_s & id_y + d_x  \end{pmatrix} \, .
\end{equation}

Under spin rotations, the singlet component $f_s$ will then transform as a scalar, while the triplet component $\underline{d}=(d_x,d_y,d_z)$ transforms as an ordinary vector.
Another useful feature of this notation is that it becomes almost trivial to distinguish between short-range and long-range triplet components; the projection $d_\pll = \v d \cdot \uv h$ along the exchange field corresponds to the SRTs, while the perpendicular part $d_\perp = \abs{\v d \times \uv h}$ describes the LRTs, where $\uv h$ here denotes the unit vector of the exchange field. 
For a concrete example, if the exchange field is oriented along the $z$-axis, then $d_z$ will be the short-range component, while both $d_x$ and $d_y$ are long-ranged components.
In the coming sections, we will demonstrate that the LRT component can be identified from its density of states signature, as measurable by tunneling spectroscopy.

In the limit of weak proximity effect, we may linearize both the Usadel equation and Kupriyanov--Lukichev boundary conditions.
Using the singlet/triplet decomposition in \eqnref{Eqn:wpegamma}, and the Rashba--Dresselhaus coupling in \eqnref{Eqn:A}, the linearized version of the Usadel equation can be written:
\begin{align}
	\frac{i}{2} D_F\partial^2_z f_s    &= \epsilon f_s    +   \v h \cdot \v d 					\label{Eqn:wpeUsadelS}\; ,\\
	\frac{i}{2} D_F\partial^2_z \v d\, &= \epsilon \v d \,+\, \v h f_s  + 2iD_Fa^2 \v{\Omega}(\chi) \, \v d	\; ,	\label{Eqn:wpeUsadel}
\end{align}
where we for brevity have defined an SO interaction matrix
\begin{align}
	\v \Omega(\chi) = \begin{pmatrix} 1 & -\sin 2\chi & 0 \\ -\sin 2\chi & 1 & 0 \\ 0 & 0 & \phantom{-}2\phantom{-} \end{pmatrix}  \; .
\end{align}
We have now condensed the Usadel equation down to two coupled differential equations for $f_s$ and $\v d$, where the coupling is proportional to the exchange field and the SO interaction term. The latter has been written as a product of a factor $2iD_Fa^2$, depending on the strength~$a$, and a factor $\v\Omega(\chi)\v d$, depending on the angle $\chi$ in the polar notation. 
The matrix $\v\Omega(\chi)$ becomes diagonal for a Dresselhaus coupling with $\chi = 0$ or a Rashba coupling with $\chi = \pm \pi/2$, which implies that there is no triplet mixing for such systems.
In contrast, the off-diagonal terms are maximal for $\chi = \pm \pi/4$, which suggests that the triplet mixing is maximal when the Rashba and Dresselhaus coefficients have the same magnitude.
In addition to the off-diagonal triplet mixing terms, we see that the diagonal terms of $\v \Omega(\chi)$ essentially result in imaginary energy contributions $2iD_Fa^2$.
As we will see later, this can in some cases result in a suppression of all the triplet components in the ferromagnet.

We will now consider exchange fields in the $xy$-plane, 
\begin{equation}
	\v h = h\,\cos \theta \; \uv x + h\,\sin \theta \; \uv y \; .
\end{equation}
Since the linearized Usadel equations show that the presence of a singlet component $f_s$ only results in the generation of triplet components along $\v h$, and the SO interaction term only mixes the triplet components in the $xy$-plane, the only nonzero triplet components will in this case be $d_x$ and $d_y$.
The SRT amplitude $d_\pll$ and LRT amplitude $d_\perp$ can therefore be written:
\begin{align}
	d_\pll  &= \phantom{-} d_x \, \cos\theta + d_y \, \sin \theta\; , \\
	d_\perp &=          -  d_x \, \sin\theta + d_y \, \cos \theta\; .
\end{align}
By projecting the linearized Usadel equation for $\v d$ along the unit vectors $(\cos \theta, \sin \theta, 0)$ and $(-\sin \theta, \cos \theta,0)$, respectively, then we obtain coupled equations for the SRTs and LRTs:
\begin{align}
	\frac{i}{2} D_F\partial^2_z f_s     =\,& \epsilon f_s    +  h d_\pll				\; ,	\label{eq:usadel-lin-s}											\\
	\frac{i}{2} D_F\partial^2_z d_\pll  =\,& [\epsilon + 2iD_Fa^2 (1-\sin 2\theta \, \sin 2\chi)] \, d_\pll \notag\\
					    &- 2iD_Fa^2 \cos 2\theta \, \sin 2\chi \;d_\perp + hf_s	\; ,          \label{eq:usadel-lin-pll}\\
	\frac{i}{2} D_F\partial^2_z d_\perp =\,& [\epsilon + 2iD_Fa^2 (1+\sin 2\theta \, \sin 2\chi)] \, d_\perp \notag\\
					    &- 2iD_Fa^2 \cos 2\theta \, \sin 2\chi \;d_\pll  		\; .              \label{eq:usadel-lin-perp}
\end{align}
These equations clearly show the interplay between the singlet component $f_s$, SRT component $d_\pll$, and LRT component $d_\perp$.
If we start with only a singlet component $f_s$, then the presence of an exchange field $h$ results in the generation of the SRT component $d_\pll$.
The presence of an SO field can then result in the generation of the LRT component $d_\perp$, where the mixing term is proportional to $a^2 \cos 2\theta \sin 2\chi$.
This implies that in the weak proximity limit, LRT mixing is absent for an exchange field direction $\theta=\pi/4$, corresponding to $h_x=h_y$, while it is maximized if $\theta=\{0,\pi/2,\pi\}$ and at the same time $\chi = \pm \pi/4$.
In other words, the requirement for maximal LRT mixing is therefore that the exchange field is aligned along either the $x$-axis or $y$-axis, while the Rashba and Dresselhaus coefficients should have the same magnitude. It is important to note here that although the mixing between the triplet components is maximal at $\theta=\{0,\pi/2,\pi\}$, this does not necessarily mean that the signature of the triplets in physical quantities is most clearly seen for these angles, as we shall discuss in detail later.

Moreover, these equations show another interesting consequence of having an SO field in the ferromagnet, which is unrelated to the LRT generation.
Note that the effective quasiparticle energies coupling to the SRTs and LRTs become
\begin{align}
	E_\pll  = \epsilon + 2iD_Fa^2 (1 - \sin2\theta \, \sin 2\chi) \; \label{eq:E-SRT},\\
	E_\perp = \epsilon + 2iD_Fa^2 (1 + \sin2\theta \, \sin 2\chi) \; \label{eq:E-LRT}.
\end{align}
When $\theta=\chi=\pm\pi/4$, then the SRTs are entirely unaffected by the presence of SO coupling; the triplet mixing term vanishes for these parameters, and $E_\pll$ is also clearly independent of $a$.
However, when $\theta = -\chi = \pm\pi/4$, the situation is drastically different.
There is still no possibility for LRT generation, however the SRT energy $E_\pll = \epsilon + 4iD_Fa^2$ will now obtain an imaginary energy contribution which destabilizes the SRTs.
In fact, numerical simulations show that this energy shift destroys the SRT components as $a$ increases.
As we will see in \secref{results-Tc}, this effect results in an increase in the critical temperature of the bilayer.
Thus, switching between $\theta = \pm \pi/4$ in a system with $\chi \simeq\pm\pi/4$ may suggest a novel method for creating a triplet spin valve.

When $\chi = \pm\pi/4$ but $\theta \neq \pm\pi/4$, the triplet mixing term proportional to $\cos 2\theta \, \sin 2\chi$ will no longer vanish, so we get LRT generation in the system.
We can then see from the effective triplet energies that as $\theta \rightarrow \mathrm{sgn}(\chi)\pi/4$, the imaginary part of $E_\pll$ vanishes, while the imaginary part of $E_\perp$ increases.
This leads to a relative increase in the amount of SRTs compared to the amount of LRTs in the system.
In contrast, as $\theta \rightarrow -\mathrm{sgn}(\chi)\pi/4$, the imaginary part of $E_\perp$ vanishes, and the imaginary part of $E_\pll$ increases.
This means that we would expect a larger LRT generation for these parameters, up until the point where the triplet mixing term $\cos 2\theta \, \sin 2\chi$ becomes so small that almost no LRTs are generated at all.
The ratio of effective energies coupling to the triplet component at the Fermi level $\epsilon = 0$ can be written as
\begin{equation}
	\label{eq:E0-SRT-LRT}
	\frac{E_\perp\!(0)}{E_\pll(0)} = \frac{1+\sin2\theta\,\sin2\chi}{1-\sin2\theta\,\sin2\chi}.
\end{equation}

\subsection{Density of states}
The density of states $D(\epsilon)$ containing all spin components can be written in terms of the Riccati matrices as
\begin{equation}
	D(\epsilon) = \mathrm{Tr} [ N(1+\gamma\tilde\gamma) ] /2 \; ,
\end{equation}
which for the case of zero energy can be written concisely in terms of the singlet component $f_s$ and triplet components $\v d$,
\begin{equation}
	\label{Eqn:DOS}
	D(0) = 1 - \abs{f_s(0)}^2/2 + \abs{\v d(0)}^2/2 \; .
\end{equation}
The singlet and triplet components are therefore directly competing to lower and raise the density of states \cite{tanaka_prl_07}.
Furthermore, since we are primarily interested in the proximity effect in the ferromagnetic film, we will begin by using the known BCS bulk solution in the superconductor,
\begin{eqnarray}\label{Eqn:gRbulk}
\p{\hat{g}}_{BCS} = 
\begin{pmatrix}
\cosh (\theta)  & \sinh (\theta) i\sigma_y e^{i\phi}\\
\sinh (\theta) i\sigma_y e^{-i\phi}& -\cosh (\theta)
\end{pmatrix} \, ,
\end{eqnarray}
where $\theta=\mathrm{atanh}(\Delta/\epsilon)$, and $\phi$ is the superconducting phase. 
Using \eqnref{Eqn:wpegamma} and the definition of the tilde operation, and comparing $\hat{g}^R$ in \eqnref{Eqn:gR} with its standard expression in a bulk superconductor \eqnref{Eqn:gRbulk}, we can see that at zero energy the singlet component $f_s(0)$ must be purely imaginary and the asymmetric triplet $d_z(0)$ must be purely real if the superconducting phase is $\phi=0$.

By inspection of \eqnref{Eqn:wpeUsadel}, we can see that a transformation $h_x \leftrightarrow h_y$ along with $d_x \leftrightarrow d_y$ leaves the equations invariant.
The density of states will therefore be unaffected by such permutations,
\begin{equation}
	D[\v h = (a,b,0)] = D[\v h=(b,a,0)] \; ,
\end{equation}
while in general 
\begin{equation}
	D[\v h = (a,0,b)] \neq D[\underline{h}=(b,0,a)] \; .
\end{equation}
However, whenever one component of the planar field is exactly twice the value of the other component, one can confirm that the linearized equations remain invariant under a rotation of the exchange field
\begin{equation}
	\v{h}=(a,2a,0) \rightarrow \v{h}=(a,0,2a) \; ,
\end{equation}
with associated invariance in the density of states.

\subsection{Critical temperature}\label{sec:theory-Tc}
When superconducting correlations leak from a superconductor and into a ferromagnet in a hybrid structure, there will also be an inverse effect, where the ferromagnet effectively drains the superconductor of its superconducting properties due to tunneling of Cooper pairs.
Physically, this effect is observable in the form of a reduction in the superconducting gap $\Delta(z)$ near the interface at all temperatures.
Furthermore, if the temperature of the hybrid structure is somewhat close to the bulk critical temperature $T_{cs}$ of the superconductor, this \emph{inverse proximity effect} can be strong enough to make the superconducting correlations vanish entirely throughout the system.
Thus, proximity-coupled hybrid structures will in practice always have a critical temperature $T_c$ that is lower than the critical temperature $T_{cs}$ of a bulk superconductor.
Depending on the exact parameters of the hybrid system, $T_c$ can sometimes be significantly smaller than $T_{cs}$, and in some cases it may even vanish ($T_c\to 0$).

To quantify this effect, it is no longer sufficient to solve the Usadel equation in the ferromagnet only.
We will now also have to solve the Usadel equation in the superconductor,
\begin{equation}
	\label{eq:usadel-s}
	D_S \partial_z^2 \gamma = -2i\epsilon \gamma - \Delta (\pauli y - \gamma \pauli y \gamma) 
				  -2(\partial_z \gamma)\tilde N\tilde \gamma(\partial_z\gamma) \; ,
\end{equation}
along with a self-consistency equation for the gap $\Delta(z)$,
\begin{equation}
	\label{eq:gap-s}
	\Delta(z) = N_0\lambda \int\limits_0^{\Delta_0 \cosh(1/N_0\lambda)\hspace{-5em}} \d\epsilon\; \re{f_s(z,\epsilon)} \tanh\left( \frac{\pi}{2e^\gamma} \frac{\epsilon/\Delta_0}{T/T_{cs}} \right) \, ,
\end{equation}
where $N_0$ is the density of states per spin at the Fermi level, and $\lambda > 0$ is the electron-electron coupling constant in the BCS theory of superconductivity.
For a derivation of the gap equation, see \appref{gap}.

To study the effects of the SO coupling on the critical temperature of an SF structure, we therefore have to find a self-consistent solution to \eqnref{Eqn:SOUsadel} in the ferromagnet, \eqnref{Eqn:KLRic} at the interface, and \eqnrefs{eq:usadel-s}{eq:gap-s} in the superconductor.
In practice, this is done by successively solving one of the equations at a time numerically, and continuing the procedure until the system converges towards a self-consistent solution.
To obtain accurate results, we typically have to solve the Usadel equation for 100--150 positions in each material, around 500 energies in the range $(0,2\Delta_0)$, and 100 more energies in the range $(2\Delta_0,\omega_c)$, where the Debye cutoff $\omega_c \approx 76\Delta_0$ for the superconductors considered herein.
This procedure will then have to be repeated up to several hundred times before we obtain a self-consistent solution for any given temperature of the system.
Furthermore, if we perform a conventional linear search for the critical temperature $T_c/T_{cs}$ in the range $(0,1)$ with a precision of 0.0001, it may require up to 10,000 such iterations to complete, which may take several days depending on the available hardware and efficiency of the implementation.
The speed of this procedure may, however, be significantly increased by performing a binary search instead.
Using this strategy, the critical temperature can be determined to a precision of $1/2^{12+1} \approx 0.0001$ after only 12 iterations, which is a significant improvement.
The convergence can be further accelerated by exploiting the fact that $\Delta(z)$ from iteration to iteration should decrease monotonically to zero if $T>T_c$; however, the details will not be further discussed in this paper.

\section{Results}\label{Sec:Results}
We consider the proximity effect in an SF bilayer in \ref{SubSec:SF}, using the BCS bulk solution for the superconductors. The case of pure Rashba coupling is discussed in \ref{SubSec:PureR}, and the SFS Josephson junction is treated in \ref{SubSec:SFS}.
We take the thin-film layering direction to be oriented in the $z$-direction and fix the spin-orbit coupling to Rashba--Dresselhaus type in the $xy$-plane as given by \eqnref{Eqn:A}.
We set $L_F/\xi_S=0.5$. The coherence length for a diffusive bulk superconductor typically lies in the range $10-30$ nm. We solve the equations using \textsc{Matlab} with the boundary value differential equation package \texttt{bvp6c} and examine the density of states $D(\epsilon)$ for energies normalised to the superconducting gap $\Delta$.
For brevity of notation, we include the normalization factor in the coefficients $\alpha$ and $\beta$ in these sections.
This normalization is taken to be the length of the ferromagnetic region $L_F$, so that for instance $\alpha=1$ in the figure legends means ${\alpha L_F=1}$.
Finally, in \secref{results-Tc}, we calculate the dependence of the critical temperature of an SF bilayer as a function of the different system parameters.

\subsection{SF Bilayer}\label{SubSec:SF}
Consider the SF bilayer depicted in Fig.~\ref{Fig:model}(a).  In section \ref{SubSec:Theory_SO} we introduced the conditions for the LRT component to appear, and from Eq. (\ref{Eqn:cond}) it is clear that no LRTs will be generated if the exchange field is aligned with the layering direction, \ie $\underline{h}\parallel\uv{z}$, since  Eq.~(\ref{Eqn:cond}) will be parallel to the exchange field. Conversely, the general condition for LRT generation with in-plane magnetisation is both that $h_x\neq h_y$ and that the SO coupling is not of pure Rashba or pure Dresselhaus form. However, it became clear in Section \ref{SubSec:WPE} that the triplet mixing was maximal for equal Rashba and Dresselhaus coupling strengths, and in fact the spectroscopic signature is quite sensitive to deviations from this. 

In Ref. \onlinecite{KontosRyazanov}, the density of states for an SF bilayer was shown to display oscillatory behavior as a function of distance penetrated into the ferromagnet. The physical origin of this stems from the non-monotonic dependence of the superconducting order parameter inside the F layer, which oscillates and leads to an alternation of dominant singlet and dominant triplet correlations as a function of distance from the interface. When the triplet ones dominate, the proximity-induced change in the density of states is inverted compared to SN structures, giving rise to an enhancement of the density of states at low-energies in this so-called $\pi$-phase where the proximity-induced superconducting order parameter is negative. 

\begin{figure}[H]
  \centering
  \includegraphics[width=\linewidth]{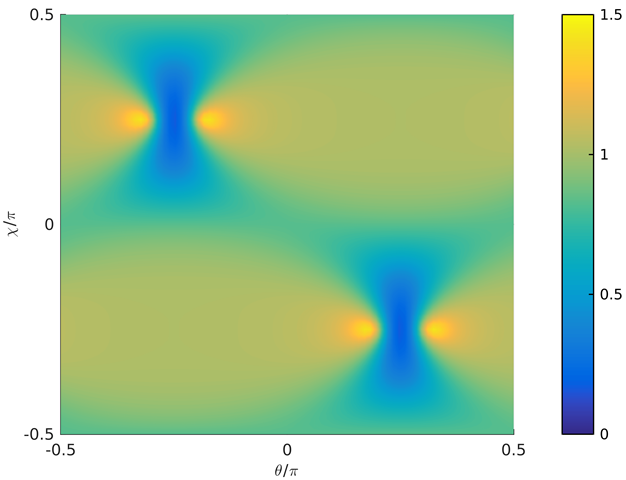}
  \caption
   {Zero-energy density of states $D(0)$ as a function of the spin-orbit angle $\chi$ and magnetization angle $\theta$. We have used a ferromagnet of length $L_F/\xi_S=0.5$ with an exchange field $h/\Delta=3$ and a spin-orbit magnitude $a\xi_S=2$.}
	\label{Fig:zepBilayer}
\end{figure}

For SF bilayers without SO coupling and a homogeneous exchange field, one expects to see a spectroscopic minigap whenever the Thouless energy is much greater than the strength of the exchange field.
The minigap in SF structures closes when the resonant condition $h \sim E_g$ is fulfilled, where $E_g$ is the minigap occuring without an exchange field, and a zero-energy peak emerges instead\cite{Yokoyama2005}. The minigap $E_g$ depends on both the Thouless energy and the resistance of the junction. For stronger fields we will have an essentially featureless density of states (see \eg \refcite{Kawabata2013} and references therein). This is indeed what we observe for $\alpha=\beta=0$ in Fig.~\ref{Tab:SFhz}. With purely out-of-plane magnetisation $\underline{h}\parallel\uv{z}$, the effect of SO coupling is irrespective of type: Rashba, Dresselhaus or both will always create a minigap. With in-plane magnetisation however, the observation of a minigap above the SO-free resonant condition $h>E_g$ indicates that dominant Rashba or dominant Dresselhaus coupling is present. The same is true for SFS trilayers, and thus to observe a signature of long-range triplets the Rashba--Dresselhaus coefficients must be similar in magnitude, and in the following we shall primarily focus on this regime. To clarify quantitatively how much the Rashba and Dresselhaus coefficients can deviate from each other before destroying the low-energy enhancement of the density of states, which is the signature of triplet Cooper pairs in this system, we have plotted in Fig. \ref{Fig:zepBilayer} the density of states at the Fermi level ($\varepsilon=0$) as a function of the spin-orbit angle $\chi$ and the magnetization direction $\theta$. For purely Rashba or Dresselhaus coupling $(\chi=\{0,\pm \pi/2\})$, the deviation from the normal-state value is small. However, as soon as both components are present a highly non-monotonic behavior is observed. This is particularly pronounced for $\chi \rightarrow \pm\pi/4$, although the conversion from dominant triplets to dominant singlets as one rotates the field by changing $\theta$ is seen to occur even away from $\chi=\pm \pi/4$.

\begin{figure}[H]
  \centering
  \includegraphics[width=\linewidth]{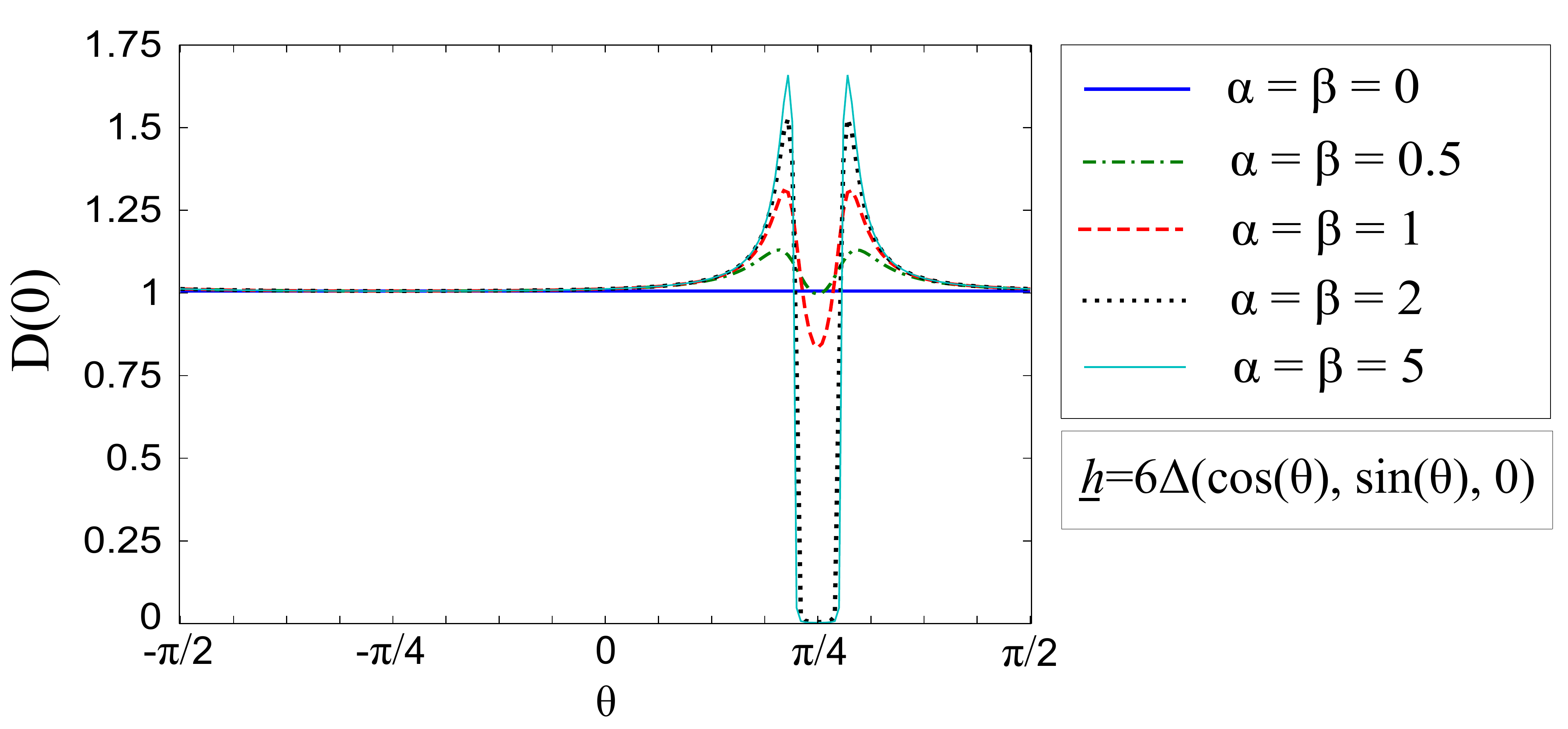}
  \caption
   {The dependence of the density of states of the SF bilayer at zero energy on the angle $\theta$ between the $x$ and $y$ components of the magnetisation exchange field $\underline{h}/\Delta=6(\cos(\theta),\sin(\theta),0)$ for increasing SO coupling. As the strength of the SO coupling increases we see increasingly sharp variations in the density of states from an optimal peak at around $\theta\approx7\pi/32$ and $\theta\approx9\pi/32$ to a gap around $\theta=\pi/4$.}
	\label{Fig:DoSAngle}
\end{figure}

With either $\underline{h}=h\underline{\hat{x}} \neq 0$, or equivalently $\underline{h}=h\underline{\hat{y}}\neq 0$, LRTs are generated provided $\alpha\beta\neq 0$, and in Fig.~\ref{Tab:SFhy} we can see that the addition of SO coupling introduces a peak in the density of states at zero energy, which saturates for a certain coupling strength. This peak manifests as sharper around $\varepsilon=0$ than the zero-energy peak associated with weak field strengths of the order of the gap (\ie as evident from $\alpha=\beta=0$ in Fig.~\ref{Tab:SFhy}), which occurs regardless of magnetisation direction or texture\cite{Kawabata2013,Yokoyama2005}. By analysing the real components of the triplets, for a gauge where the superconducting phase is zero, we can confirm that this zero-energy peak is due to the LRT component, in this case $d_x$, also depicted in Fig.~\ref{Tab:SFhy}, in agreement with the predictions for \textit{textured} magnetisation \textit{without} SO coupling \cite{Kawabata2013}.  However, it is also evident from Fig.~\ref{Tab:SFhy} that  increasing the field strength rapidly suppresses the density of states towards that of the normal metal, making the effect more difficult to detect experimentally. The way to ameliorate this situation is to remember that the introduction of SO coupling means the \textit{direction} of the exchange field is crucially important, as we see in Fig.~\ref{Fig:DoSAngle}, and this allows for a dramatic spectroscopic signature for fields without full alignment with the $x$- or $y$-axes.

Fig.~\ref{Fig:DoSAngle} shows how the density of states at zero energy varies with the angle $\theta$ between $h_x$ and $h_y$ at zero energy; with $\theta=0$ the field is aligned with $h_x$, and with $\theta=\pi/2$ it is aligned with $h_y$. We see that the inclusion of SO coupling introduces a nonmonotonic angular dependance in the density of states, with increasingly sharp features as the SO coupling strength increases, although the optimal angle at approximately $\theta=7\pi/32$ and $\theta=9\pi/32$ varies minimally with increasing SO coupling.
Clearly the ability to extract maximum LRT conversion from the inclusion of SO coupling is highly sensitive to the rotation angle, with near step-function behaviour delineating the regions of optimal peak in the density of states and an energy gap for strong SO coupling.
It is remarkable to see how $D(0)$ vs. $\theta$ formally bears a strong resemblance to the evolution of a fully gapped BCS \cite{bcs} density of states $D(\varepsilon)$ vs. $\varepsilon$ to a flat density of states as the SO coupling decreases.

These results can again be explained physically by the linearized equations \eqref{eq:usadel-lin-s}--\eqref{eq:usadel-lin-perp}.
Since the case $\alpha=\beta$ corresponds to $\chi=-\pi/4$ in the notation developed in the preceding sections, \eqnref{eq:E0-SRT-LRT} implies that $E_\perp(0)>E_\pll(0)$ when $\theta < 0$, while $E_\perp(0)<E_\pll(0)$ when $\theta>0$.
In other words, for negative $\theta$, the SO coupling suppresses the LRT components, and the exchange field suppresses the other components.
Since the singlet and SRT components have opposite sign in \eqnref{Eqn:DOS}, this renders the density of states essentially featureless.
However, for positive $\theta$, both the SO coupling and the exchange field suppress the SRT components, meaning that LRT generation is energetically favoured.
Note that $E_\perp/E_\pll \rightarrow \infty$ as $\theta \rightarrow +\pi/4$, which explains why the LRT generation is maximized in this regime.
Since the triplet mixing term in \eqnref{eq:usadel-lin-perp} is proportional to $(\cos 2\theta \; \sin 2\chi)$, the LRT component vanishes when the value of $\theta$ gets too close to $+\pi/4$.
Furthermore, since $E_\pll$ has a large imaginary energy contribution in this case, the SRTs are also suppressed at $\theta = +\pi/4$.
Thus, despite LRTs being most energetically favored at this exact point, we end up with a system dominated by singlets due to the SRT suppression and lack of LRT production pathway. Nevertheless, one would conventionally expect that exchange fields of a magnitude $h\gg \Delta$ as depicted in Fig.\ref{Fig:DoSAngle} would suppress any features in the density of states, while we observe an obvious minigap. \textit{Thus, the singlet correlations become much more resilient against the pair-breaking effect of the exchange field when spin-orbit coupling is present.}

To identify the physical origin of this effect, we solve the linearized equations \eqref{eq:usadel-lin-s}--\eqref{eq:usadel-lin-perp} along with their corresponding boundary conditions for the specific case $\varepsilon=0$, $\theta=-\chi=\pi/4$. We consider a bulk superconductor occupying the space $x<0$ while the ferromagnet length $L_F$ is so large that one in practice only needs to keep the decaying parts of the anomalous Green's function. We then find the following expression for the singlet component at the SF interface in the absence of SO coupling:
\begin{align}
f_s^0 = \frac{\sinh (\textrm{arctanh}(\Delta/\epsilon))}{2\zeta L_F}\sqrt{\frac{D_F}{h}}.
\end{align}
With increasing $h$, the singlet correlations are suppressed in the conventional manner. However, we now incorporate SO coupling in the problem. For more transparent analytical results, we focus on the case $2(a\xi)^2 \gg h/\Delta$. This condition can be rewritten as $2D_F a^2 \gg h$. In this case, a similar calculation gives the singlet component at the SF interface in the presence of SO coupling:
\begin{align}
f_s = f_s^0 \sqrt{\frac{D_F a^2}{2h}}.
\end{align}
Clearly, the SO coupling enhances the singlet component in spite the presence of an exchange field since $\sqrt{{D_F a^2}/{h}}\gg1$. This explains the presence of the conventional zero energy gap for large SO coupling even with a strong exchange field. A consequence of this observation is that SO coupling in fact provides a route to a \textit{magnetically tunable minigap}. Fig. \ref{Fig:DoSAngle} shows that when both an exchange field and SO coupling is present, the direction of the field determines when a minigap appears. This holds even for strong exchange fields $h \gg \Delta$ as long as the SO coupling is sufficiently large as well.

We recall that the LRT Cooper pairs, defined as the components of $\underline{d}$ perpendicular to $\underline{h}$, may be characterized by a quantity $d_\perp$ which is defined by the cross product of the two vectors: $d_\perp = |\underline{d}\times \underline{\hat{h}}|$. We saw above that the spectroscopic signature of LRT generation is strongly dependent on the angle of the field, and this angle is a tunable parameter for sufficiently weak magnetic anisotropy. In Fig.~\ref{Tab:SFhxhy} we see an example of the effect this rotation can have on the spectroscopic signature of LRT generation: when the exchange field is changed from $\underline{h}=(6\Delta,3\Delta,0)\rightarrow (6\Delta,5\Delta,0)$, \ie changing the direction of the field, we see that a strong zero-energy peak emerges due to the presence of LRT in the system. This large peak emerges despite the stronger exchange field that would ordinarily reduce the density of states towards the normal state, \ie as in Fig.~\ref{Tab:SFhy} for $\underline{h}=\Delta\underline{\hat{y}}\rightarrow 3\Delta\underline{\hat{y}}$. If one were to remove the SO coupling, the low-energy density of states would thus have no trace of any superconducting proximity effect, which demonstrates the important role played by the SO interactions here. Finally, for completeness we include an example of the effect of rotating the field to have a component along the junction in Fig.~\ref{Tab:SFhyhz}. Comparing the case of  $\underline{h}=(0,3\Delta, 6\Delta)$ in Fig.~\ref{Tab:SFhyhz} with $\underline{h}=(6\Delta,3\Delta,0)$ in Fig.~\ref{Tab:SFhxhy}, we see that the two cases are identical, as predicted in the limit of weak proximity effect, and increasing the magnitude of the out-of-plane $z$ component of the field has no effect on the height of the zero-energy peak, which is instead governed by the in-plane $y$ component.

\newpage
\begin{widetext}
\begin{minipage}{0.97\textwidth}
\begin{figure}[H]
  \centering
  \includegraphics[width=\textwidth]{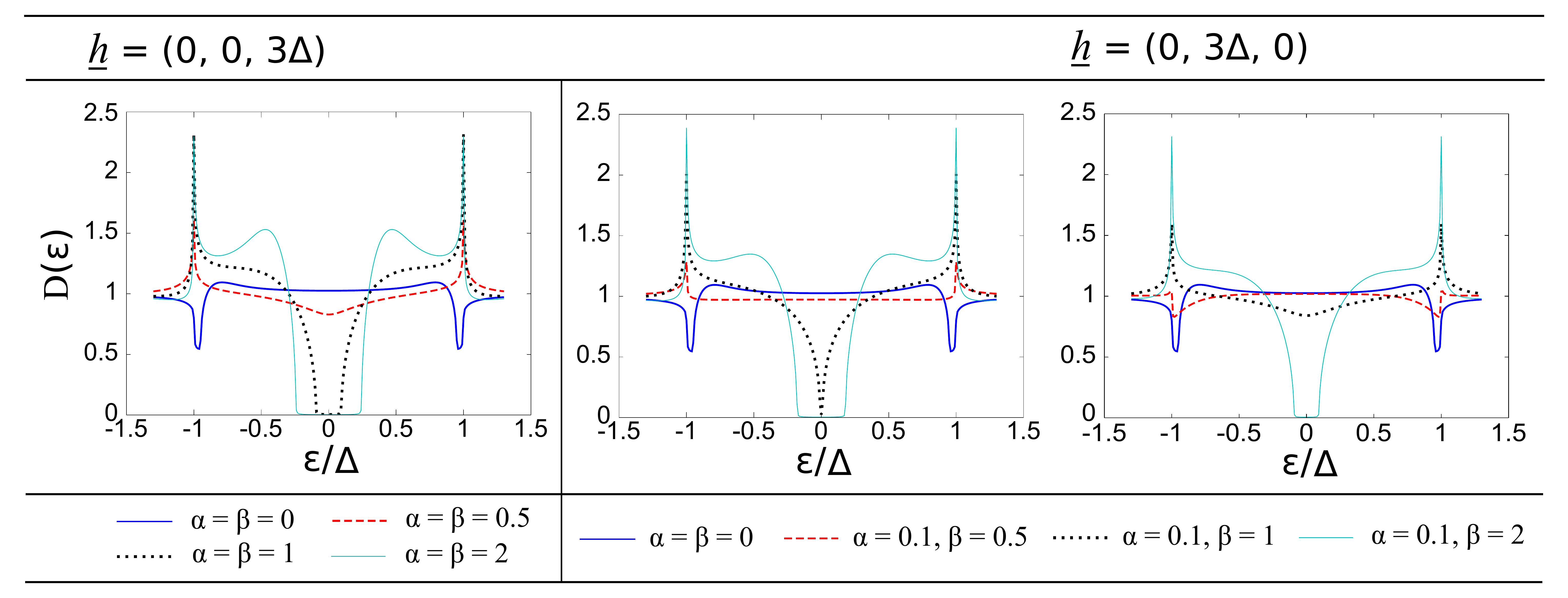}
	\begin{minipage}{0.975\textwidth}
  \caption
   {Density of states $D(\epsilon)$ for the SF bilayer with energies normalised to the superconducting gap $\Delta$ and SO coupling normalised to the inverse ferromagnet length $1/L_F$. The table shows the spectroscopic effect of increasing SO coupling with $\alpha=\beta$ when the magnetisation $\underline{h}=3\Delta\hat{z}$, \ie with the field perpendicular to the interface, and the effect of increasing difference between the Rashba and Dresselhaus coefficients for both $\underline{h}=3\Delta\hat{z}$ and $\underline{h}=3\Delta\hat{y}$. Although the conditions for LRT generation are fulfilled in the latter case, it is clear that no spectroscopic signature of this is present.}
	\label{Tab:SFhz}
	\end{minipage}
\end{figure}
\begin{figure}[H]
  \centering
  \includegraphics[width=\textwidth]{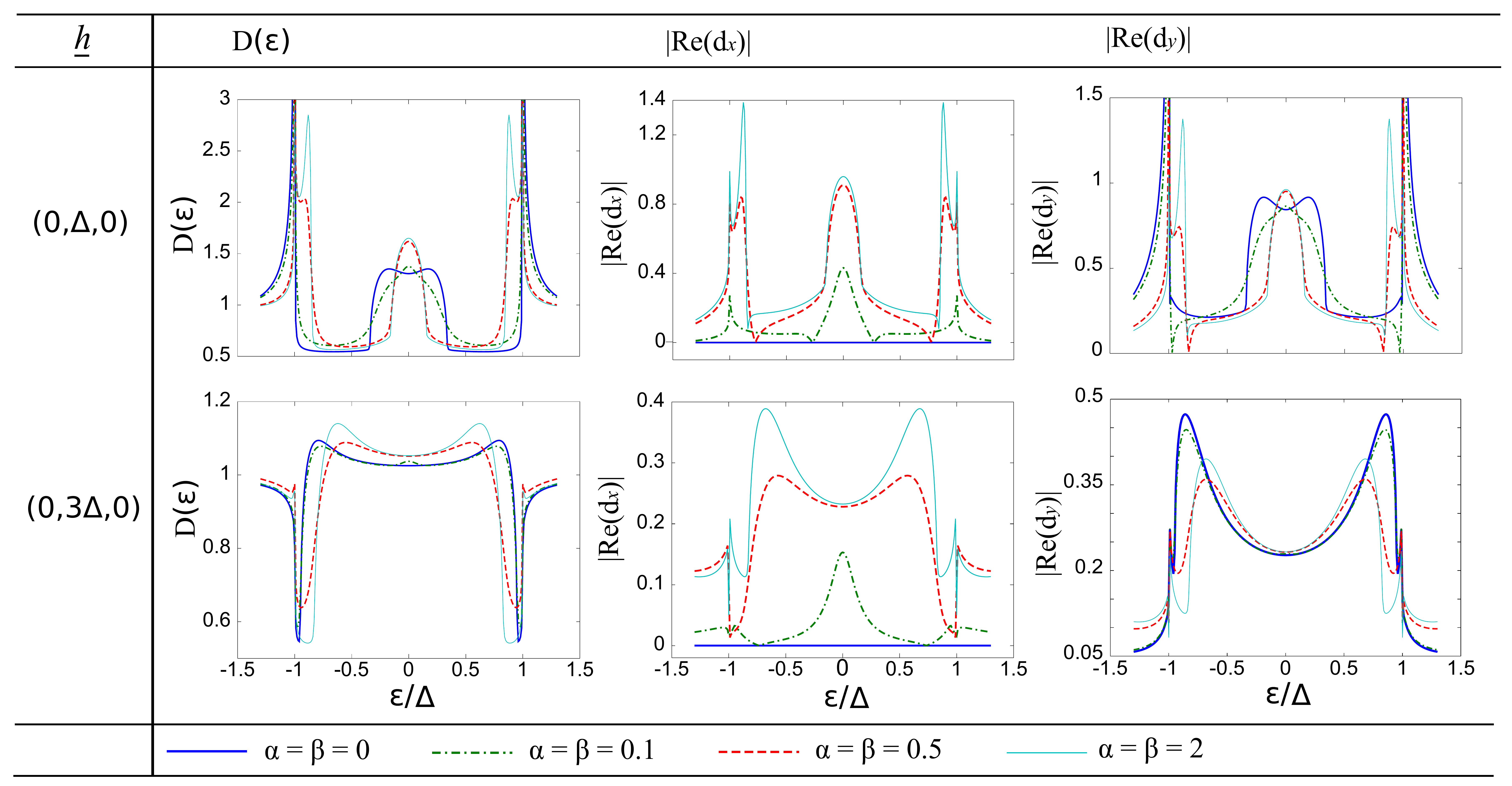}
	\begin{minipage}{0.975\textwidth}
  \caption
   {Density of states $D(\epsilon)$ for the SF bilayer with energies normalised to the superconducting gap $\Delta$ and SO coupling normalised to the inverse ferromagnet length $1/L_F$. The table shows the spectroscopic effect of equal Rashba--Dresselhaus coefficients when the magnetisation is oriented entirely in the $y$-direction, and also the correlation between the SO-induced zero-energy peak with the long-range triplet component $|\mathrm{Re}(d_x)|\equiv \mathrm{Re}(d_\perp)$. It is clear that the predominant effect of the LRT component, which appears only when the SO coupling is included, is to increase the peak at zero energies. Increasing the field strength rapidly suppresses the density of states towards that of the normal metal.}
		\label{Tab:SFhy}
	\end{minipage}
\end{figure}
\end{minipage}
\clearpage
\end{widetext}
\begin{widetext} 
\begin{minipage}{0.97\textwidth}
\begin{figure}[H]
  \centering
  \includegraphics[width=\textwidth]{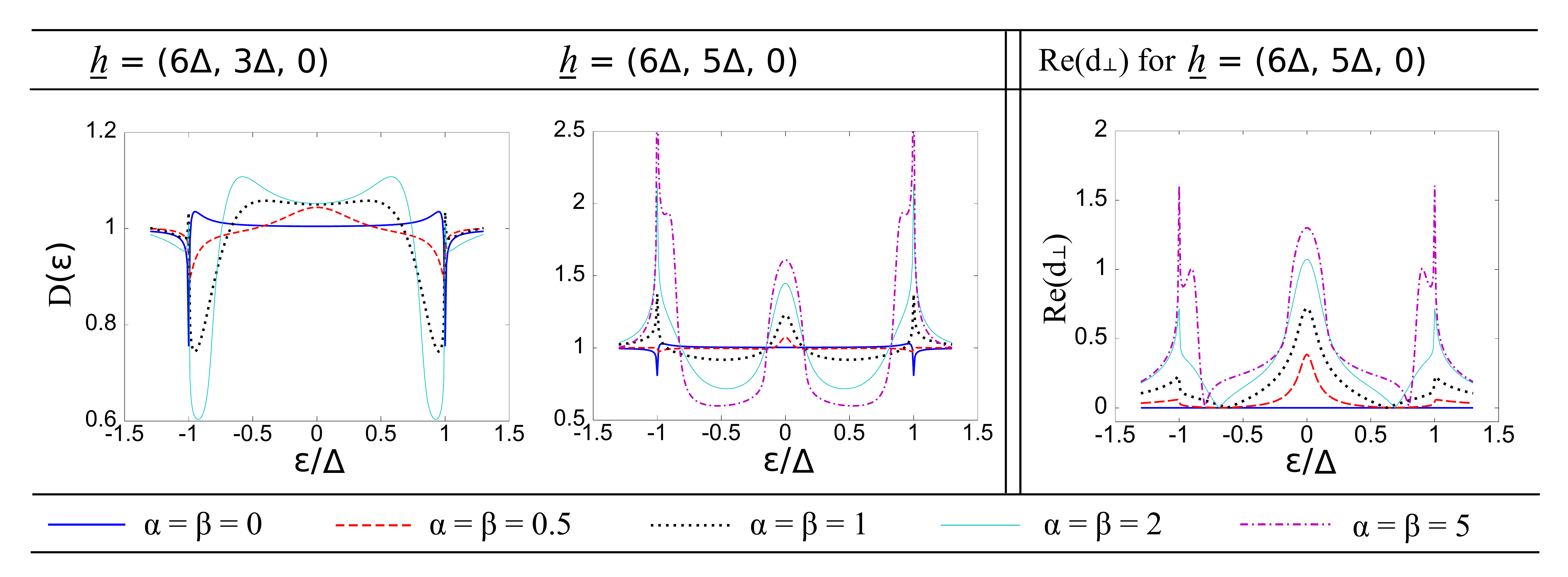}
  \begin{minipage}{0.965\textwidth}
  \caption
   {Density of states $D(\epsilon)$ in the SF bilayer for energies normalised to the superconducting gap $\Delta$ and SO coupling normalised to the inverse ferromagnet length $1/L_F$. The table shows the spectroscopic features of the SF bilayer with rotated exchange field in the $xy$-plane. Again we see a peak in the density of states at zero energy due to the LRT component, \ie the component of $\underline{d}$ perpendicular to $\underline{h}$, $d_\perp$. The height of this zero-energy peak is strongly dependent on the angle of the field vector in the plane, as shown in Fig.~\ref{Fig:DoSAngle}. For near-optimal field orientations increasing the SO coupling leads to a dramatic increase in the peak of the density of states at zero energy.}
		\label{Tab:SFhxhy}
  \end{minipage}
\end{figure}
\begin{figure}[H]
  \centering
  \includegraphics[width=\textwidth]{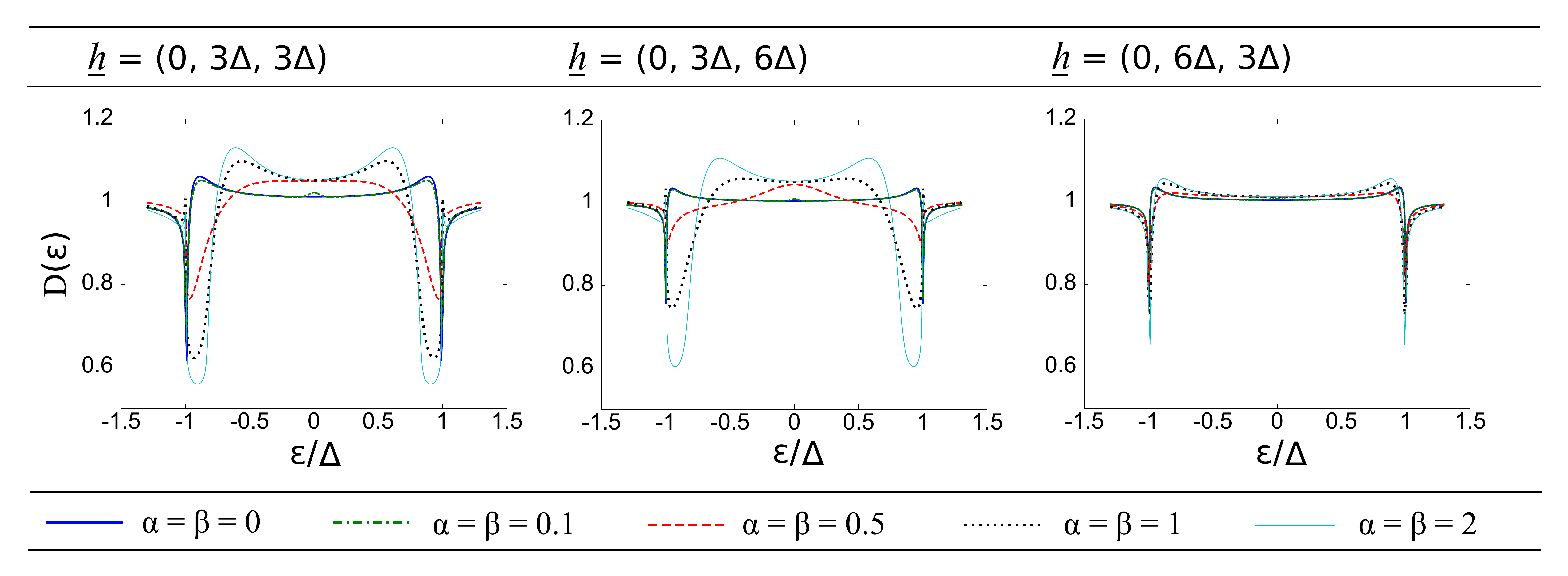}
  \begin{minipage}{0.965\textwidth}
  \caption
   {Density of states $D(\epsilon)$ in the SF bilayer for energies normalised to the superconducting gap $\Delta$ and SO coupling normalised to the inverse ferromagnet length $1/L_F$. The table shows the spectroscopic features of the SF bilayer with a rotated exchange field in the $xz\equiv yz$-plane. Note that when the field component along the junction is twice the component in the $y$-direction, here $\underline{h}=(0,3\Delta,6\Delta)$, the density of states is equivalent to the case $\underline{h}=(6\Delta,3\Delta,0)$ illustrated in Fig.~\ref{Tab:SFhxhy}, as predicted in the limit of weak proximity effect.}
		\label{Tab:SFhyhz}
  \end{minipage}
\end{figure}
\end{minipage}
\clearpage
\end{widetext}

\subsection{SF bilayer with pure Rashba coupling}\label{SubSec:PureR}

There exists another experimentally viable setup where the LRT can be created.
In the case where pure Rashba SO coupling is present, originating e.g. from interfacial
asymmetry, the condition for the existence of LRT is that the exchange field has a 
component both in-plane and out-of-plane. Although the LRT formally is non-zero,
it is desirable to clarify if and how it can be detected through spectroscopic signatures.

From an experimental point of view, it is known that PdNi and CuNi \cite{birge} can in general feature a canted magnetization orientation relative to the film-plane due to the competition between shape anisotropy and magnetocrystalline anisotropy. This is precisely the situation required in order to have an exchange field with both an in-plane ($xy$-plane in our notation) and out-of-plane ($z$-direction) component. In our model, the ferromagnetism coexists with the Rashba SO coupling, which may be taken as a simplified
model of two separate layers where the SO coupling is induced e.g. by a very thin heavy metal and PdNi or CuNi is deposited on top of it.

\begin{figure}[H]
  \centering
  \includegraphics[width=\linewidth]{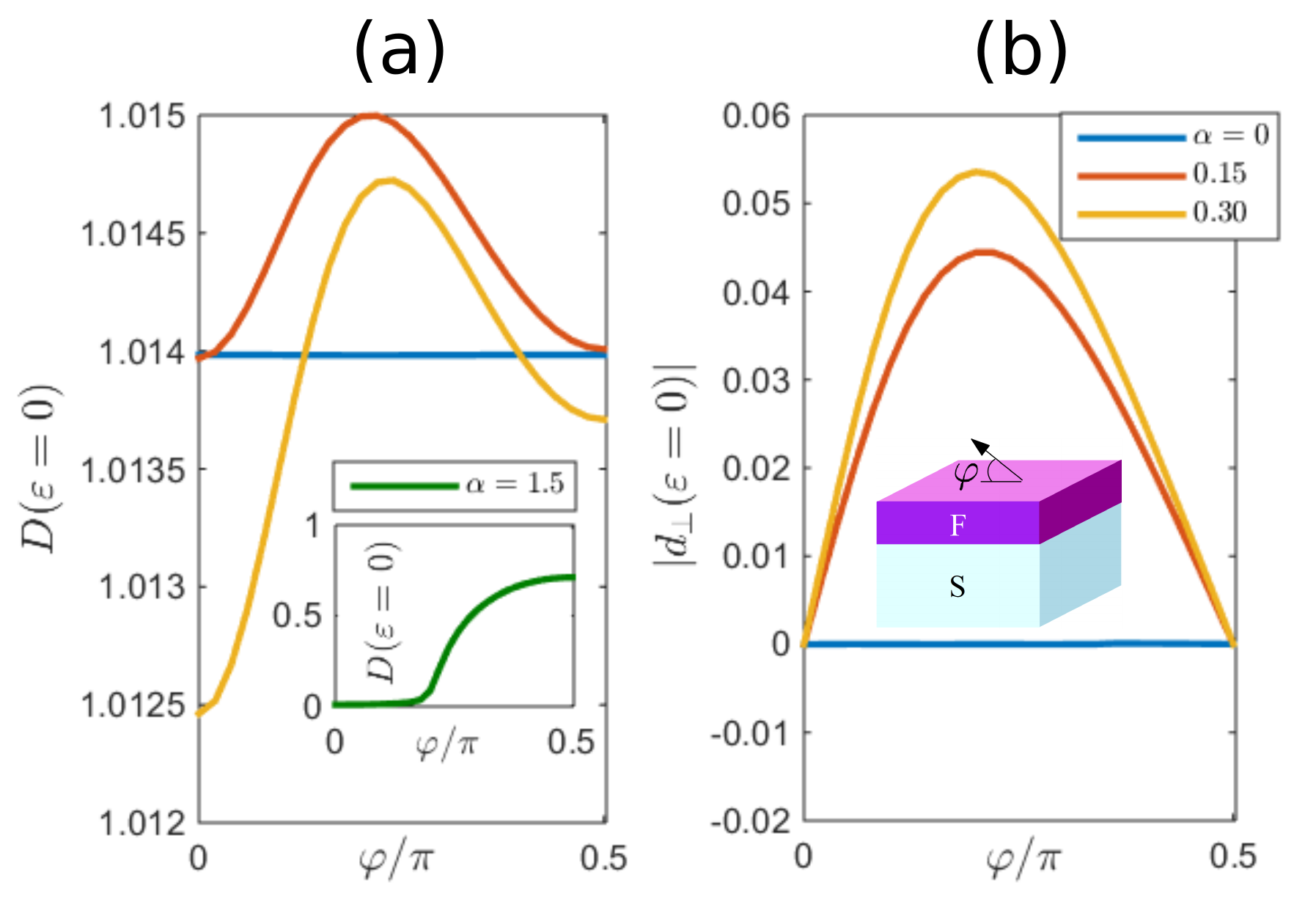}
  \caption
   {(Color online) (a) Plot of the zero-energy density of states $D(0)$ in an S/F structure with pure Rashba spin-orbit coupling. We have set $h/\Delta=4$ and $L/\xi_S=0.5$. Inset: stronger SO coupling $\alpha=1.5$, demonstrating that the angular variation of $D(0)$ remains, although the enhancement due to triplets is absent. (b) Plot of the magnitude of the LRT anomalous Green's function $|d_\perp|$ at $\varepsilon=0$. As seen, its enhancement correlates with an accompanying increase in the density of states
for the same angle $\varphi$, and beyond an optimal SO coupling value there is anticorrelation between the density of states peak and $|d_\perp|$. The only angle of importance is the angle $\varphi$ between the out-of-plane and in-plane component of the exchange field, shown in the inset.}
\label{Fig:PureR}
\end{figure}

To determine how the low-energy density of states is influenced by the triplet pairing, we plot in Fig. \ref{Fig:PureR}(a) $D(0)$ as a function of the misalignment angle $\varphi$ between the film-plane and its perpendicular axis [see inset of Fig. \ref{Fig:PureR}(b) for junction geometry]. In order to correlate the spectroscopic features with the LRT, we plot in Fig. \ref{Fig:PureR}(b) the LRT Green's function $|d_\perp|$. It is clear that the LRT vanishes when $\varphi=0$ or $\varphi=\pi/2$. This is consistent with the fact that for pure Rashba coupling, purely in-plane or out-of-plane direction of the exchange field gives $d_\perp=0$
according to our previous analysis.  However, for $\varphi \in (0,\pi/2)$ the LRT exists. Its influence on $D(0)$ is seen in Fig. \ref{Fig:PureR}(a): an enhancement of the zero-energy density of states. For any particular set of junction parameters there is an optimal value of the SO coupling, and in approaching this value the density of states is correlated with Re$\{d_\perp\}$. Beyond this optimal value, they are anticorrelated, as evident from Fig. \ref{Fig:PureR} as the SO coupling increases, but the angular correlation remains. We note that the magnitude of the enhancement of the density of states is substantially smaller than what we obtained with both Rashba and Dresselhaus coupling. At the same time, the magnitude of the enhancement is of precisely the same order as previous experimental works that have measured the density of states in S/F structures \cite{KontosRyazanov, SanGiorgio2008}. 

Note that it is only the angle between the plane and the tunneling direction which is of importance: the density of states is invariant under a rotation in the film-plane of the exchange field. The SO-induced enhancement of the zero-energy density of states reaches an optimal peak before further increases in the magnitude of the Rashba coupling results in a suppression of both the short- and long-ranged triplet components, causing the low-energy density of states enhancement to vanish. The correlation with the LRT component $|d_\perp|$ correspondingly changes to anticorrelation, evident in Fig. \ref{Fig:PureR}.
Nevertheless, the strong angular variation with $D(0)$ remains although $D(0)<1$ for all $\varphi$ [see inset of Fig. \ref{Fig:PureR}(a)]. Increasing the exchange field $h$ further suppressed the proximity effect overall. 

The main effect of the SO coupling is that $D(0)$ depends on the exchange field direction.
As seen for the case of $\alpha=0$ in Fig. \ref{Fig:PureR}(a), there is no directional dependence without SO coupling. Thus, depending on the exchange field angle between the in-plane and out-of plane direction, measuring an enhanced $D(0)$ at low-energies is a signature of the presence of LRT Cooper pairs in the ferromagnet. More generally, measuring a dependence on the exchange field direction $\varphi$ would be a direct consequence of the presence of SO coupling in the system, even in the regime of \textit{e.g.} moderate to strong Rashba coupling where the triplets are suppressed.

\subsection{Josephson junction}\label{SubSec:SFS}
By adding a superconducting region to the right interface of the SF bilayer we form an SFS Josephson junction. It is well known that the phase difference between the superconducting regions governs how much current can flow through the junction\cite{Tinkham}, and the density of states for a diffusive SNS junction has been measured experimentally with extremely high precision\cite{leSueuretal2008}. Here we consider such a transversal junction structure as depicted in Fig.\ref{Fig:model}(b), again with intrinsic SO coupling in the $xy$-plane (Eq. \ref{Eqn:A}) in the ferromagnet and with BCS bulk values for each superconductor. In \ref{SubSec:Junc1} we consider single orientations along the principal axes of the system $(x,y,z)$ of the uniform exchange field and in \ref{SubSec:JuncRot} we consider a rotated field. Experimentally, the density of states can be probed at the superconductor/ferromagnet interface if one of the superconductors is a superconducting island, and the scanning tunneling microscope approaches from the top, next to this superconductor island.

Let us first recapitulate some known results. We saw in Section~\ref{Sec:Theory} that the spin-singlet, SRT and LRT components compete to raise and lower the density of states at low energies. Their relative magnitude is affected by the magnitude and direction of both the exchange field and SO coupling and results in three distinctive qualitative profiles: the zero-energy peak from the LRTs, the singlet-dominated regime with a minigap, and the flat, featureless profile in the absence of superconducting correlations. In the Josephson junction, the spectroscopic features are in addition sensitive to the phase difference $\phi$ between the superconductors. In junctions with an interstitial normal metal, the gap decreases as $\phi=0\rightarrow\pi$, closing entirely at $\phi=\pi$ such that the density of states is that of the isolated normal metal; identically one\cite{Zhouetal1998, leSueuretal2008}. Without an exchange field the density of states is unaffected by the SO coupling. This is because without an exchange field the equations governing the singlet and triplet components are decoupled and thus no singlet-triplet conversion can occur. From a symmetry point of view, it is reasonable that the time-reversal invariant spin-orbit coupling does not alter the singlet correlations.

Without SO coupling and as long as the exchange field is not too large, changing the phase difference can qualitatively alter the density of states from minigap to peak at zero energy (see Fig.~\ref{Tab:SFS1}), a useful feature permitting external control of the quasiparticle current flowing through the junction. The underlying reason is that the phase difference controls the relative ratio of the singlet and triplet correlations: when the singlets dominate, a minigap is induced which mirrors their origin in the bulk superconductor.

As in the bilayer case, there is a resonant condition\cite{Yokoyama2005,Kawabata2013}indicating an exchange field strength beyond which the minigap can no longer be sustained and increasing the phase difference simply lowers the density of states towards that of the normal metal. Amongst the features we outline in the following subsections, one of the effects of adding SO coupling is to make this useful gap-to-peak effect accessible with stronger exchange fields, \ie for a greater range of materials. At the same time, the SO coupling cannot be \textit{too strong} since the triplet correlations are suppressed in this regime leaving only the minigap and destroying the capability for qualitative change in the spectroscopic features.

\subsubsection{Josephson junction with uniform exchange field in single direction}\label{SubSec:Junc1}
Consider first the case in which the exchange field is aligned in a single direction, meaning that we only consider an exchange field purely along the principal $\{x,y,z\}$ axes of the system. If we again restrict the form of the SO-vector to (\ref{Eqn:A}), aligning $\underline{h}$ in the $z$-direction will not result in any LRTs. In this case the spectroscopic effect of the SO coupling is dictated by the singlet and short-range triplet features, much as in the SF bilayer case (Fig.~\ref{Tab:SFhz}). This is demonstrated in Fig.~\ref{Tab:SFS1}, where again we see a qualitative change in the density of states as the exchange field increases, with the regions of minigap and zero-energy-peak separated by the resonant condition $h \sim E_g$ without SO coupling.

We will now examine the effect of increasing the exchange field aligned in the $x$- or, equivalently, the $y$-direction. In this case, we have generation of LRT Cooper pairs. If $\underline{h}$ is sufficiently weak to sustain a gap independently of SO coupling, introducing weak SO coupling will increase the gap at zero phase difference while maintaining a peak at zero energy for a phase difference of $0.75\pi$ (see Fig.~\ref{Tab:SFS1}). Increasing the SO coupling increases this peak at zero energy up to a saturation point. As the exchange field increases sufficiently beyond the resonant condition to keep the gap closed, increasing the SO coupling increases the zero-energy peak at all phases, again due to the LRT component, eventually also reaching a saturation point. As the phase difference $\phi=0\rightarrow \pi$, the density of states reduces towards that of the normal metal, closing entirely at $\phi=\pi$ as expected\cite{Zhouetal1998,hammer_prb_07,JacobsenLinder2015}. As the value of the density of states at zero energy saturates for increasing SO coupling, fixed phase differences yield the same drop at zero energy regardless of the strength of SO coupling.

We note in passing that when the SO coupling field has a component along the junction direction $(z)$, it can qualitatively influence the nature of the superconducting proximity effect.
As very recently shown in \refcite{JacobsenLinder2015}, a giant triplet proximity effect develops at $\phi=\pi$ in this case, in complete contrast to the normal scenario of a vanishing proximity effect in $\pi$-biased junctions.

\subsubsection{Josephson junction with rotated exchange field}\label{SubSec:JuncRot}
With two components of the field $\underline{h}$, \eg from rotation, it is again useful to separate the cases with and without a component along the junction direction. When the exchange field lies in-plane (the $xy$-plane),  and provided we satisfy the conditions $h_x\neq h_y$ and $\alpha\beta\neq 0$, increasing the SO coupling drastically increases the zero energy peak as shown in Fig.~\ref{Tab:SFSrot}, again due to the LRT component. This is consistent with the bilayer behavior, where the maximal generation of LRT Cooper pairs occurs at an angle $0<\theta<\pi/4$. As the phase difference approaches $\pi$, the proximity-induced features are suppressed in the centre of the junction. This can be understood intuitively as a consequence of the order parameter averaging to zero since it is positive in one superconductor and negative in the other.

\begin{widetext}
\begin{minipage}{0.97\textwidth}
\centering
\begin{figure}[H]
  \includegraphics[width=\textwidth]{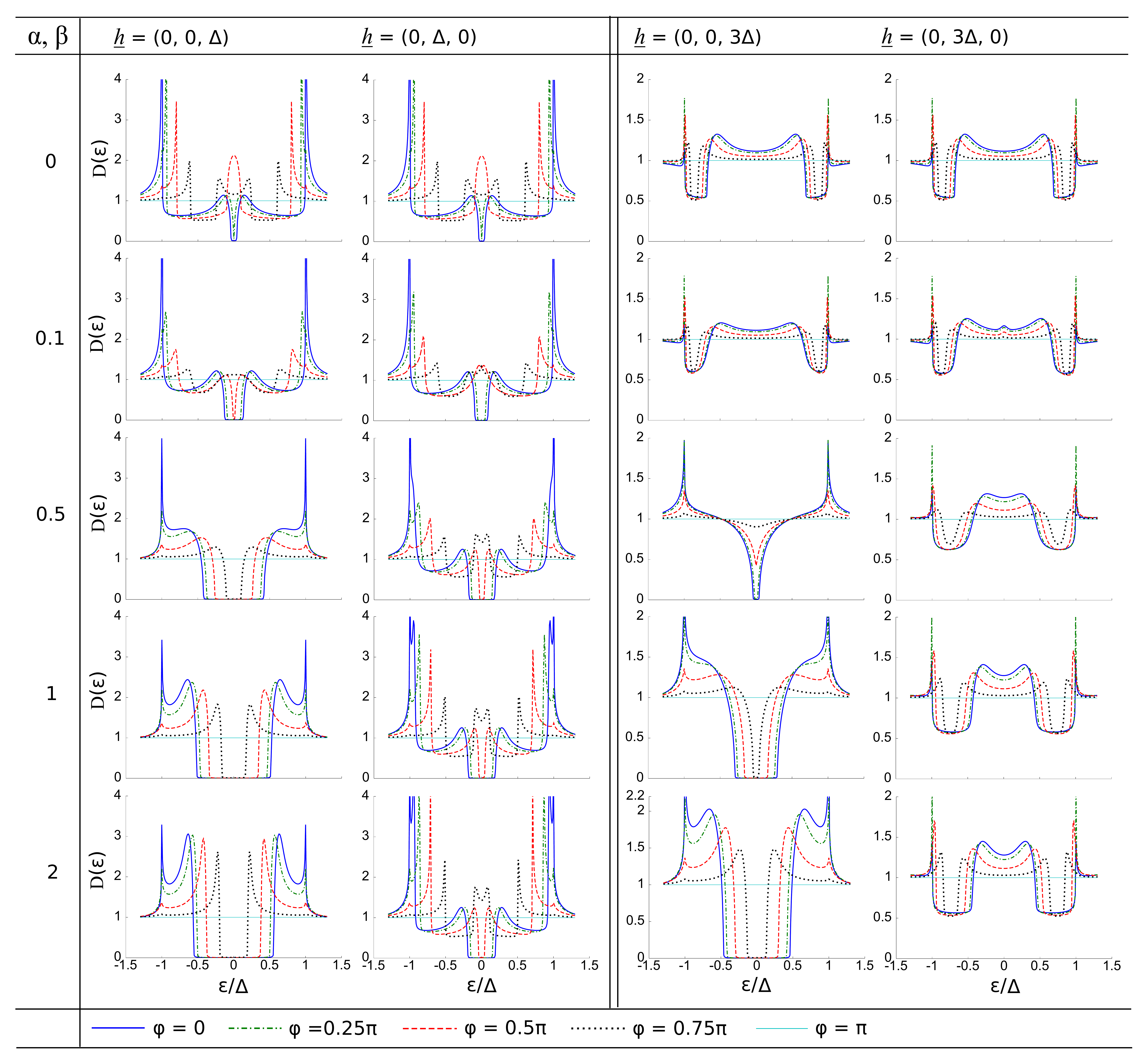}
  \begin{minipage}{0.975\textwidth}
  \caption
   {The table shows the density of states $D(\epsilon)$ in the SFS junction with increasing SO coupling and exchange field in a single direction, with $D(\epsilon)$ normalised to the superconducting gap $\Delta$ and SO coupling normalised to the inverse ferromagnet length $1/L_F$. With no SO coupling and very weak exchange field we see a phase-dictated gap-to-peak qualitative change in the density of states at zero energy. When the field is strong enough to destroy this gap, \ie above the resonant condition, increasing the phase difference simply lowers the density of states towards that of the normal metal, which is achieved at a phase difference of $\phi=\pi$. With the addition of SO coupling we see a clear difference in the density of states due to the long range triplet component, which is present when the field is oriented in $y$ but not in $z$. When LRTs are present with weak exchange fields, a phase-dictated gap-to-peak feature is retained and increased as the strength of SO coupling increases the gap, with the peak shown here at a phase difference of $0.75\pi$. For stronger exchange fields, increasing the SO coupling produces the minigap when there is no LRT component, whereas the existence of an LRT component again introduces an increasing peak at zero energy when no minigap is present.}
	\label{Tab:SFS1}
  \end{minipage}
\end{figure}
\end{minipage}

\begin{minipage}{0.97\textwidth}
\begin{figure}[H]
  \centering
  \includegraphics[width=\textwidth]{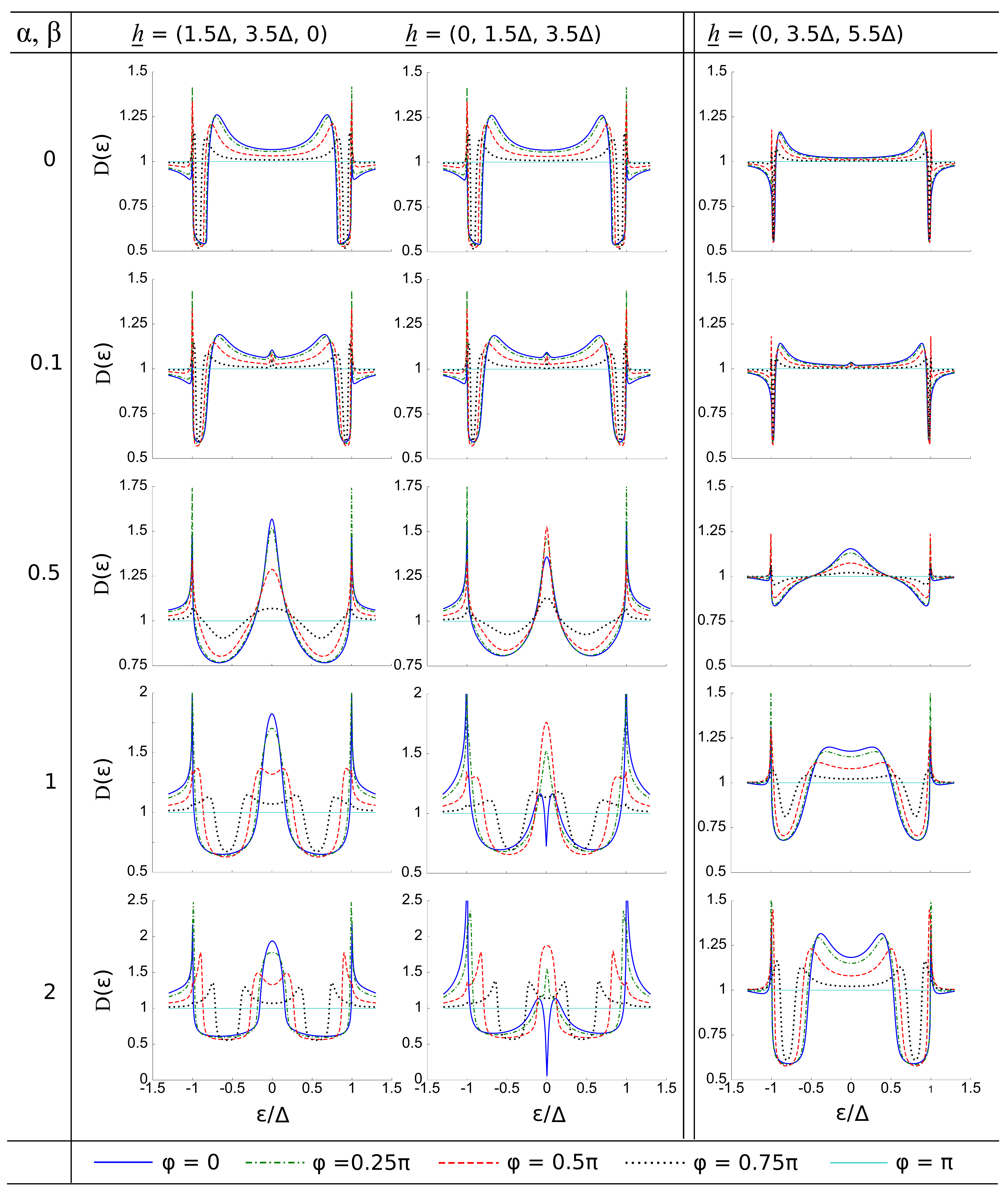}
  \begin{minipage}{0.975\textwidth}
  \caption
   {Density of states $D(\epsilon)$ in the SFS junction for energies normalised to the superconducting gap $\Delta$ and SO coupling normalised to the inverse ferromagnet length $1/L_F$. The table shows the spectroscopic effects of increasing SO coupling in SFS with rotated exchange field. In the absence of SO coupling, the density of states is flat and featureless at low energies. Increasing the SO coupling again leads to a strong increase in the peak of the density of states at zero energy, while increasing the phase difference reduces the peak and shifts the density of states weight toward the gap edge for higher SO coupling strengths. With a component of the field in the junction direction a qualitative change in the density of states from strongly suppressed to enhanced at zero energy can be achieved by altering the phase difference between the superconductors. This change can occur in the presence of stronger exchange fields when SO coupling is included. Increasing the exchange field destroys the ability to maintain a gap in the density of states and the LRT component of the SO coupling increases the zero-energy peak as it did in the bilayer case.}
	\label{Tab:SFSrot}
  \end{minipage}
\end{figure}
\end{minipage}
\end{widetext}

\begin{figure}[H]
  \centering
  \includegraphics[width=\linewidth]{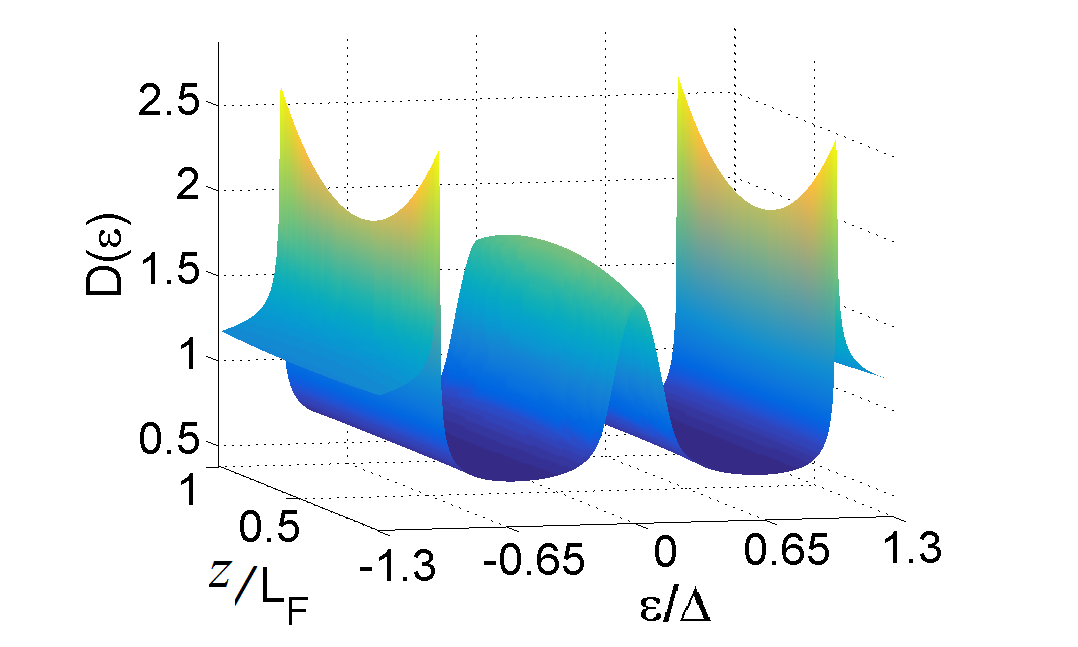}
  \caption
   {Spatial distribution of the density of states $D(\varepsilon)$ throughout the ferromagnet of an SFS junction with phase difference ${\phi=0}$, spin-orbit coupling $\alpha =\beta =1$ and magnetisation $\underline{h}=(1.5\Delta,3.5\Delta,0)$.}
	\label{Fig:SFSDoS_spatial}
\end{figure}

The 2D plots in this paper of the local density of states are given for the centre of the junction ($z=0$), where one naturally expects the relative proportion of LRTs to be greatest. However, it is interesting to note that the large peak at zero energy -- the signature of the LRTs -- is maintained throughout the ferromagnet. This is shown in Fig. \ref{Fig:SFSDoS_spatial}, for the case $\alpha =\beta =1$ and $\underline{h}=(1.5\Delta,3.5\Delta,0)$, where the maximal peak for $\phi=0$ is almost twice the normal-state value. In comparison, the depletion of this peak is surprisingly small at the superconductor interfaces.

With one component of the exchange field along the junction and another along either $x$ or $y$, a phase-dictated gap-to-peak transition at zero energy is possible with stronger fields than with the field aligned in a single direction, as shown in Fig.~\ref{Tab:SFSrot}.
Notice that in this case increasing the phase difference $\phi=0\rightarrow 0.5\pi$ gives an increase in the peak at zero energy before reducing towards the normal metal state.
For higher field strengths we find once again that increasing the SO coupling increases the peak at zero energy, up to a system-specific threshold, and increasing phase difference reduces the density of states towards that of the normal metal.

\begin{figure}[H]
  \centering
  \begin{tabular}{cc}
  \includegraphics[width=0.88\linewidth]{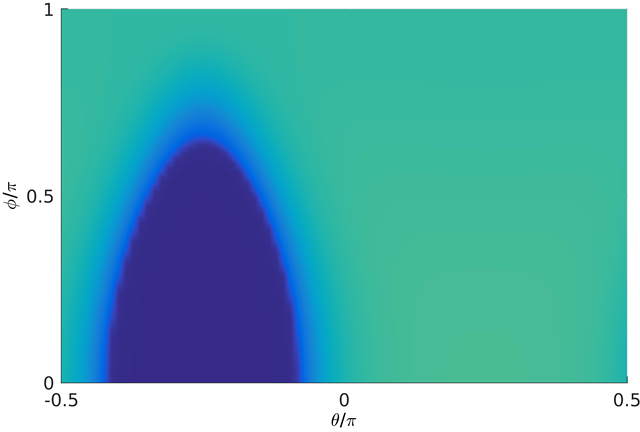} & \includegraphics[width=0.088\linewidth]{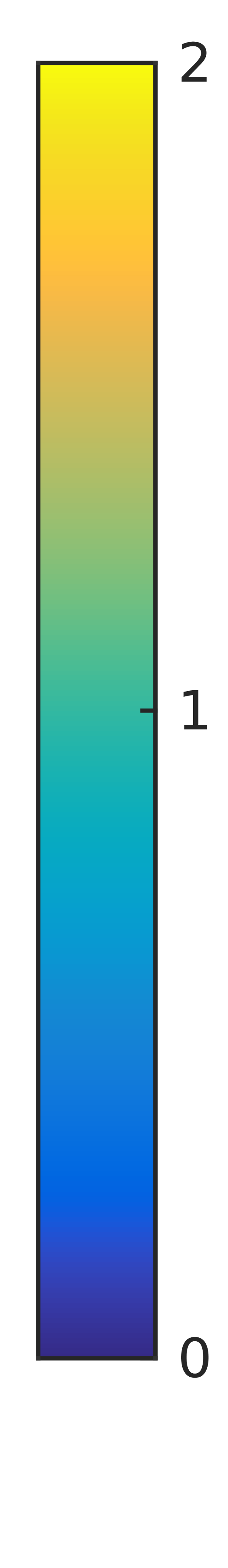}\\
  \includegraphics[width=0.88\linewidth]{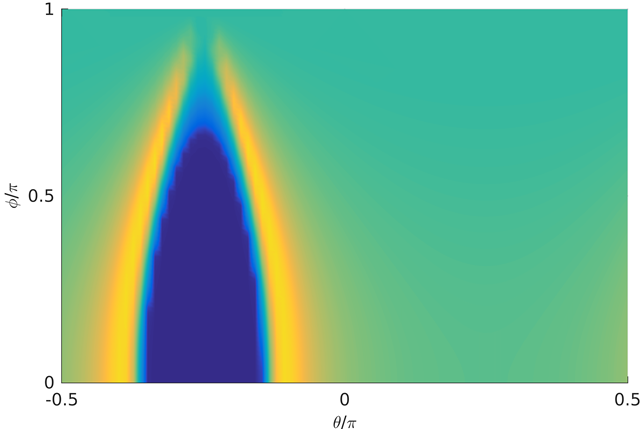}     & \includegraphics[width=0.088\linewidth]{colorbar.png}\\
  \end{tabular}
  \caption
   {Zero-energy density of states $D(0)$ as a function of the phase-difference $\phi$ and magnetization angle $\theta$, both tunable parameters experimentally. The other parameters used are $L_F/\xi_S=0.5$, $h/\Delta_0=3$, $a\xi_S=2$. In the top panel, we have dominant Dresselhaus strength ($\chi=0.15\pi$) while in the bottom panel we have equal magnitude of Rashba and Dresselhaus ($\chi=\pi/4$).}
	\label{Fig:zepJosephson}
\end{figure}

It is also useful to consider how the zero-energy density of states depends simultaneously on the phase-difference and magnetization orientation. To this end, we show in Fig. \ref{Fig:zepJosephson} a contour plot of the density of states at the Fermi level $(\varepsilon=0)$ as a function of the superconducting phase difference $\phi$ across the junction and the magnetization direction $\theta$. The proximity effect vanishes in the centre of the junction at $\phi=\pi$ for any value of the exchange field orientation, giving the normal-state value. Just as in the bilayer case (Fig. \ref{Fig:zepBilayer}), we see that the proximity effect is strongly suppressed for the range of angles $\theta>0$. When rotating the field in the opposite direction, $\theta<0$, strongly non-monotonic behavior emerges. For zero phase-difference, the physics is qualitatively similar to the bilayer situation. In this case, we proved analytically that the LRT is not produced at all when $\theta= -\pi/4$. Accordingly, Fig. \ref{Fig:zepJosephson} shows a full minigap there.

Whether or not a clear zero-energy peak can be seen due to the LRT depends on the relative strength of the Rashba and Dresselhaus coupling. In the top panel, we have dominant Dresselhaus coupling in which case the low-energy density of states show either normal-state behavior or a minigap. Interestingly, we see that the same opportunity appears in the present case of a Josephson setup as in the bilayer case: a magnetically tunable minigap appears. This effect exists as long as the phase difference is not too close to $\pi$, in which case the minigap closes. In the bottom panel corresponding to equal magnitude of Rashba and Dresselhaus, however, a strong zero-energy enhancement due to long-range triplets emerges as one moves away from $\theta=-\pi/4$. With increasing phase difference, the singlets are seen to be more strongly suppressed than the triplet correlations since the minigap region (dark blue) vanishes shortly after $\phi/\pi\simeq 0.6$ while the peaks due to triplets remain for larger phase differences.

\clearpage
\subsection{Critical temperature}
\label{sec:results-Tc}
In this section, we present numerical results for the critical temperature $T_c$ of an SF bilayer.
The theory behind these investigations is summarized in \secref{theory-Tc}, and discussed in more detail in \appref{gap}.
An overview of the physical system is given in Fig.~\ref{Fig:model}(a).
In all of the simulations we performed, we used the material parameter $N_0\lambda = 0.2$ for the superconductor, the exchange field $h = 10\Delta_0$ for the ferromagnet, and the interface parameter $\zeta = 3$ for both materials.
The other physical parameters are expressed in a dimensionless form, with lengths measured relative to the superconducting correlation length~$\xi_S$, energies measured relative to the bulk zero-temperature gap~$\Delta_0$, and temperatures measured relative to the bulk critical temperature $T_{cs}$.
This includes the SO coupling strength~$a$, which is expressed in the dimensionless form $a\xi_S$.
The plots presented in this subsection were generated from 12--36 data points per curve, where each data point has a numerical precision of 0.0001 in $T_c/T_{cs}$.
The results were smoothed with a LOESS algorithm.

\begin{figure}[H]
	\centering
	\hspace{2.5em}
	\includegraphics{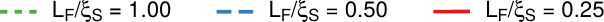}
	\hspace{-2em}
	\includegraphics[width=85mm]{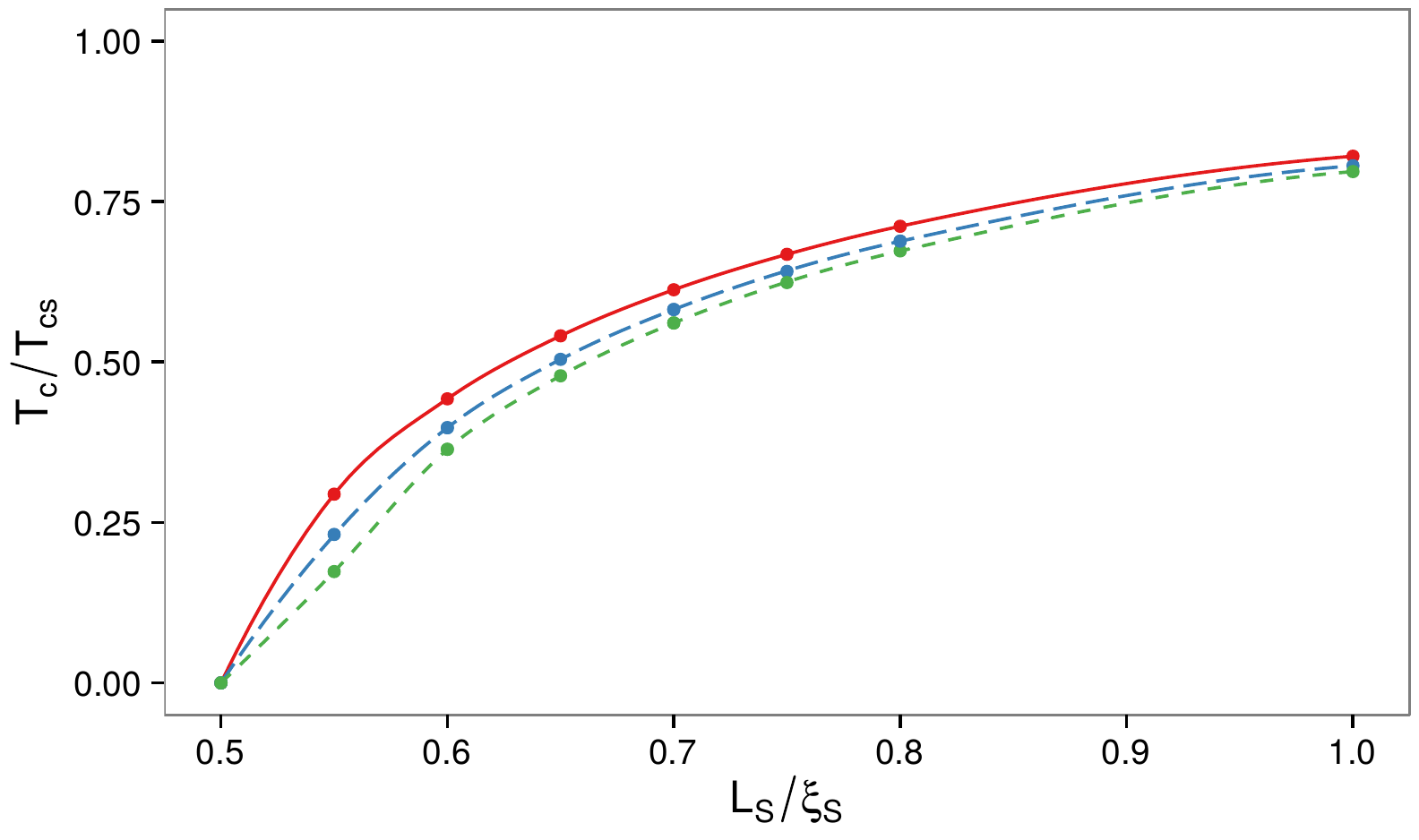}
	\caption{Plot of the critical temperature $T_c/T_{cs}$ as a function of the length $L_S/\xi_S$ of the superconductor for $a\xi_S=0$. Below a critical length $L_S$, superconductivity can no longer be sustained and $T_c$ becomes zero. For larger thicknesses of the superconducting layer, $T_c$ reverts back to its bulk value.}
	\label{fig:Tc-ds}
\end{figure}

Before we present the results with SO coupling, we will briefly investigate the effects of the ferromagnet length $L_F$ and superconductor length $L_S$ on the critical temperature, in order to identify the interesting parameter regimes.
The critical temperature as a function of the size of the superconductor is shown in \figref{Tc-ds}.
\noindent
First of all, we see that the critical temperature drops to zero when $L_S/\xi_S \approx 0.5$.
This observation is hardly surprising; since the superconducting correlation length is $\xi_S$, the critical temperature is rapidly suppressed once the length of the junction goes below $\xi_S$.
After this, the critical temperature increases quickly, already reaching nearly 50\% of the bulk value when $L_S/\xi_S = 0.6$, demonstrating that the superconductivity of the system is clearly very sensitive to small changes in parameters for this region.

\begin{figure}[H]
	\centering
	\hspace{2.5em}
	\includegraphics{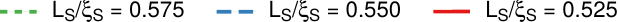}
	\hspace{-2em}
	\includegraphics[width=85mm]{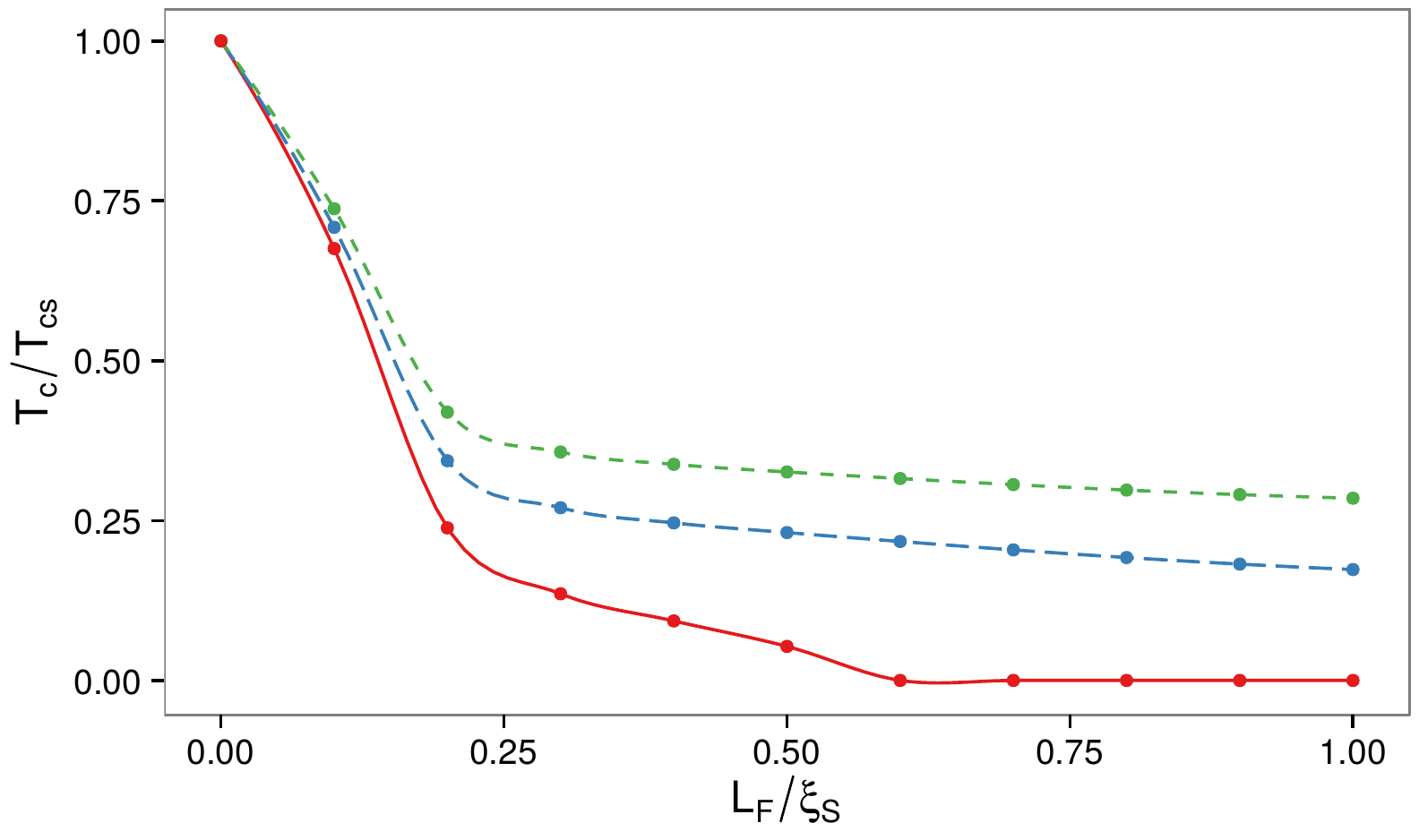}
	\caption{Plot of the critical temperature $T_c/T_{cs}$ as a function of the ferromagnet length $L_F/\xi_S$ for $a\xi_S=0$. Increasing the thickness of the ferromagnet gradually suppresses the $T_c$ of the superconductor, causing a stronger inverse proximity effect. }
	\label{fig:Tc-df}
\end{figure}

\newpage
The next step is then to observe how the behaviour of the system varies with the size of the ferromagnet, and these results are presented in Fig. \ref{fig:Tc-df}.
\noindent
We again observe that the critical temperature increases with the size of the superconductor, and decreases with the size of the ferromagnet.
The critical temperature for a superconductor with $L_S/\xi_S = 0.525$ drops to zero at $L_F/\xi_S \approx 0.6$, and stays that way as the size of the ferromagnet increases.
Thus we do not observe any strongly nonmonotonic behaviour, such as reentrant superconductivity, for our choice of parameters.
This is consistent with the results of Fominov \etal, who only reported such behaviour for systems where either the interface parameter or the exchange field is drastically smaller than for the bilayers considered herein\cite{Fominov2002}.

\begin{figure}[H]
	\centering
	\hspace{2.5em}
	\includegraphics{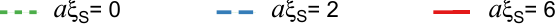}
	\hspace{-2em}
	\includegraphics[width=85mm]{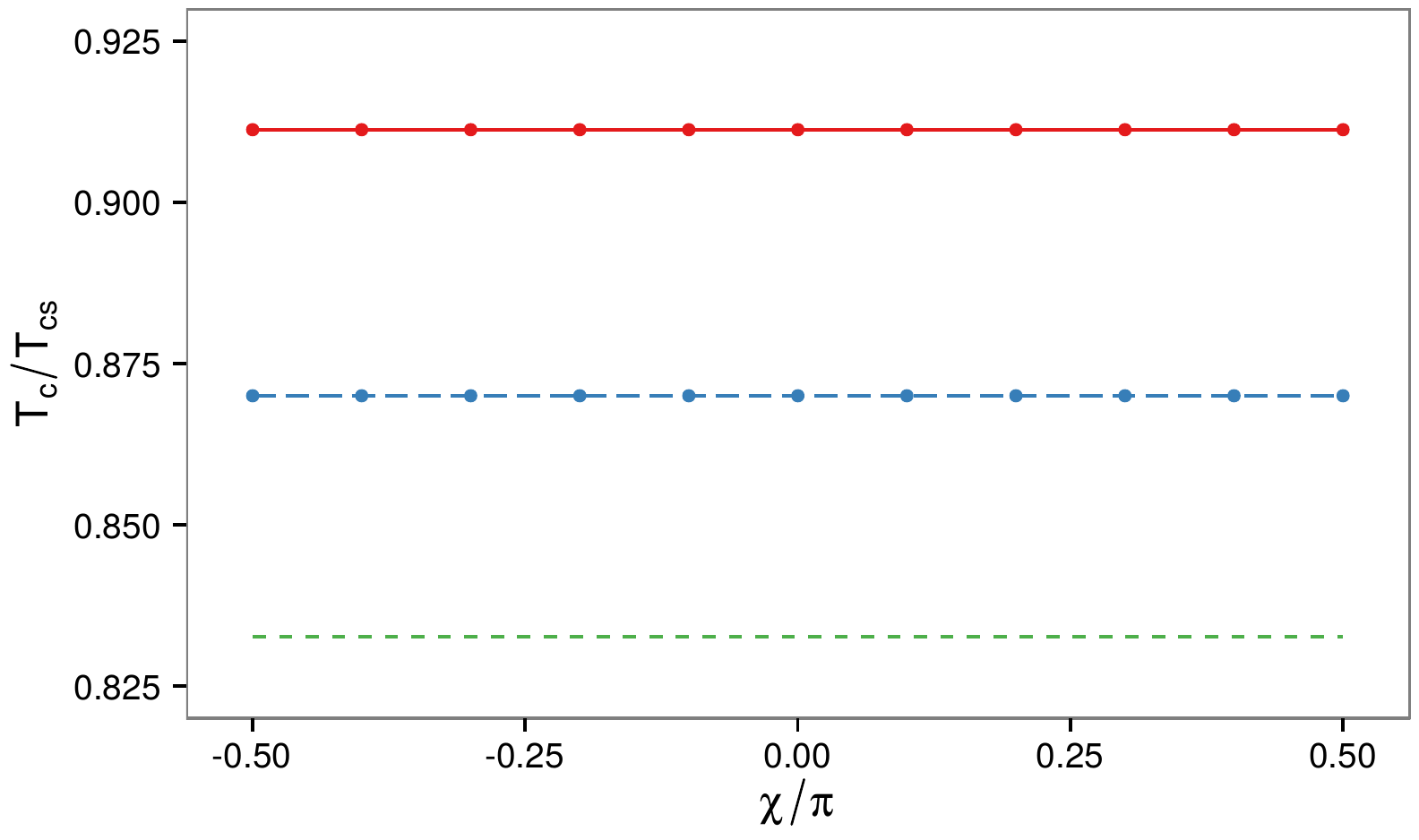}
	\caption{Plot of the critical temperature $T_c/T_{cs}$ as a function of the SO angle $\chi$, when $L_S/\xi_S = 1.00$, $L_F/\xi_S = 0.2$, and $\v h \parallel \uv z$. Increasing the SO coupling causes $T_c$ to move closer to its bulk value, since the triplet proximity effect channel becomes suppressed.}
	\label{fig:Tc-chi-z-long}
\end{figure}

We now turn to the effects of the antisymmetric SO coupling on the critical temperature, which has not been studied before.
\figrefs{Tc-chi-z-long}{Tc-chi-z-short} show plots of the critical temperature as a function of the SO angle $\chi$ for an exchange field in the $z$-direction.
The critical temperature is here independent of the SO angle $\chi$.
This result is reasonable, since the SO coupling is in the $xy$-plane, which is perpendicular to the exchange field for this geometry.
We also observe a noticeable increase in critical temperature for larger values of $a$.
This behaviour can be explained using the linearized Usadel equation.
According to \eqnref{Eqn:wpeUsadel}, the effective energy $E_z$ coupling to the triplet component in the $z$-direction becomes
\begin{equation}
	\label{eq:Ez-im}
	E_z = \epsilon + 4iD_Fa^2 \; ;
\end{equation}
so in other words, the SRTs obtain an imaginary energy shift proportional to $a^2$.
However, as shown in \eqnref{Eqn:wpeUsadelS}, there is no corresponding shift in the energy of the singlet component.
This effect reduces the triplet components relative to the singlet component in the ferromagnet, and as the triplet proximity channel is suppressed the critical temperature becomes restored to higher values.
\newpage
\begin{figure}[H]
	\centering
	\hspace{2.5em}
	\includegraphics{Tc_chi_labs.pdf}
	\hspace{-2em}
	\includegraphics[width=85mm]{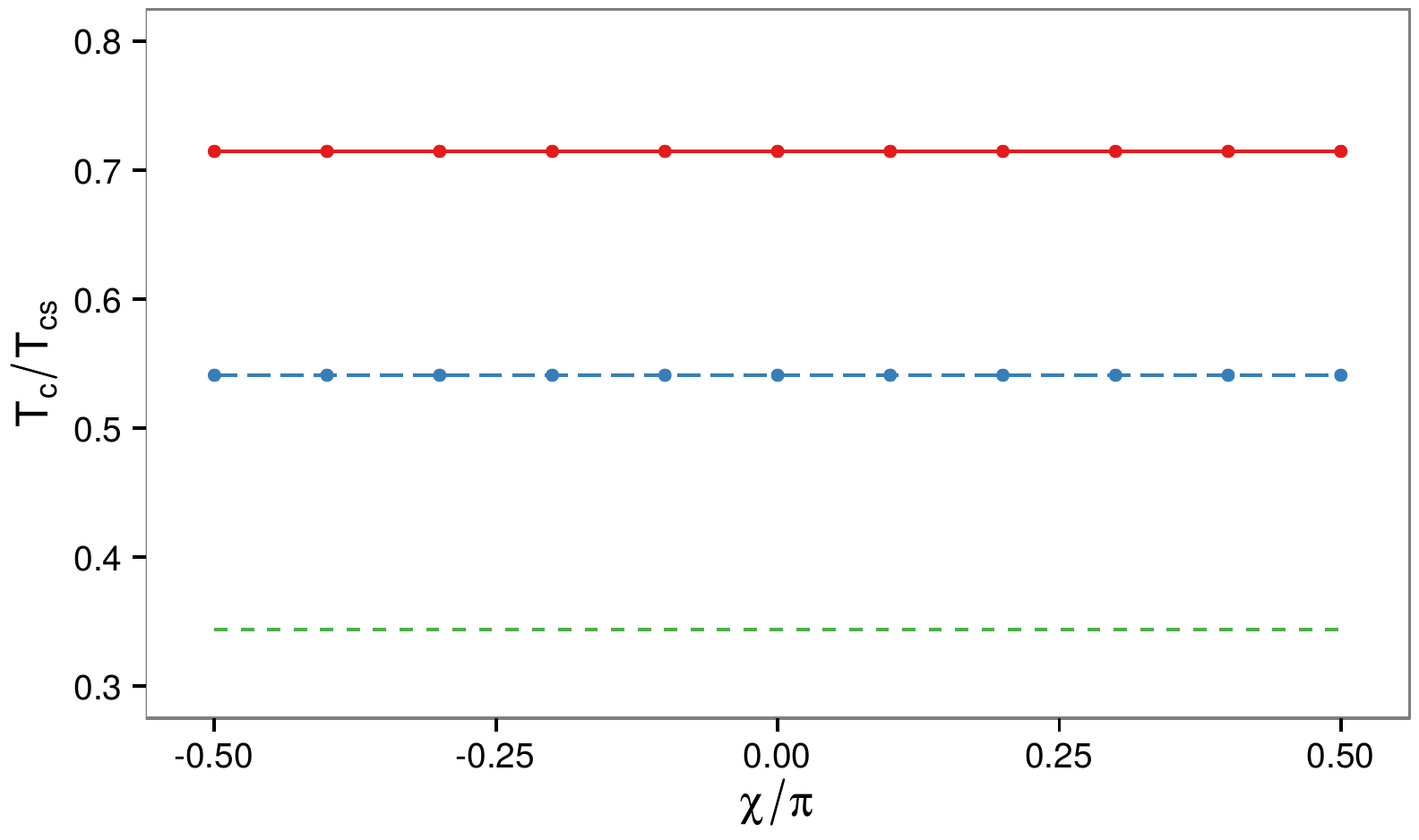}
	\caption{Plot of the critical temperature $T_c/T_{cs}$ as a function of the SO angle $\chi$, when $L_S/\xi_S = 0.55$, $L_F/\xi_S = 0.2$, and $\v h \parallel \uv z$.}
	\label{fig:Tc-chi-z-short}
\end{figure}

\begin{figure}[H]
	\centering
	\hspace{2.5em}
	\includegraphics{Tc_chi_labs.pdf}
	\hspace{-2em}
	\includegraphics[width=85mm]{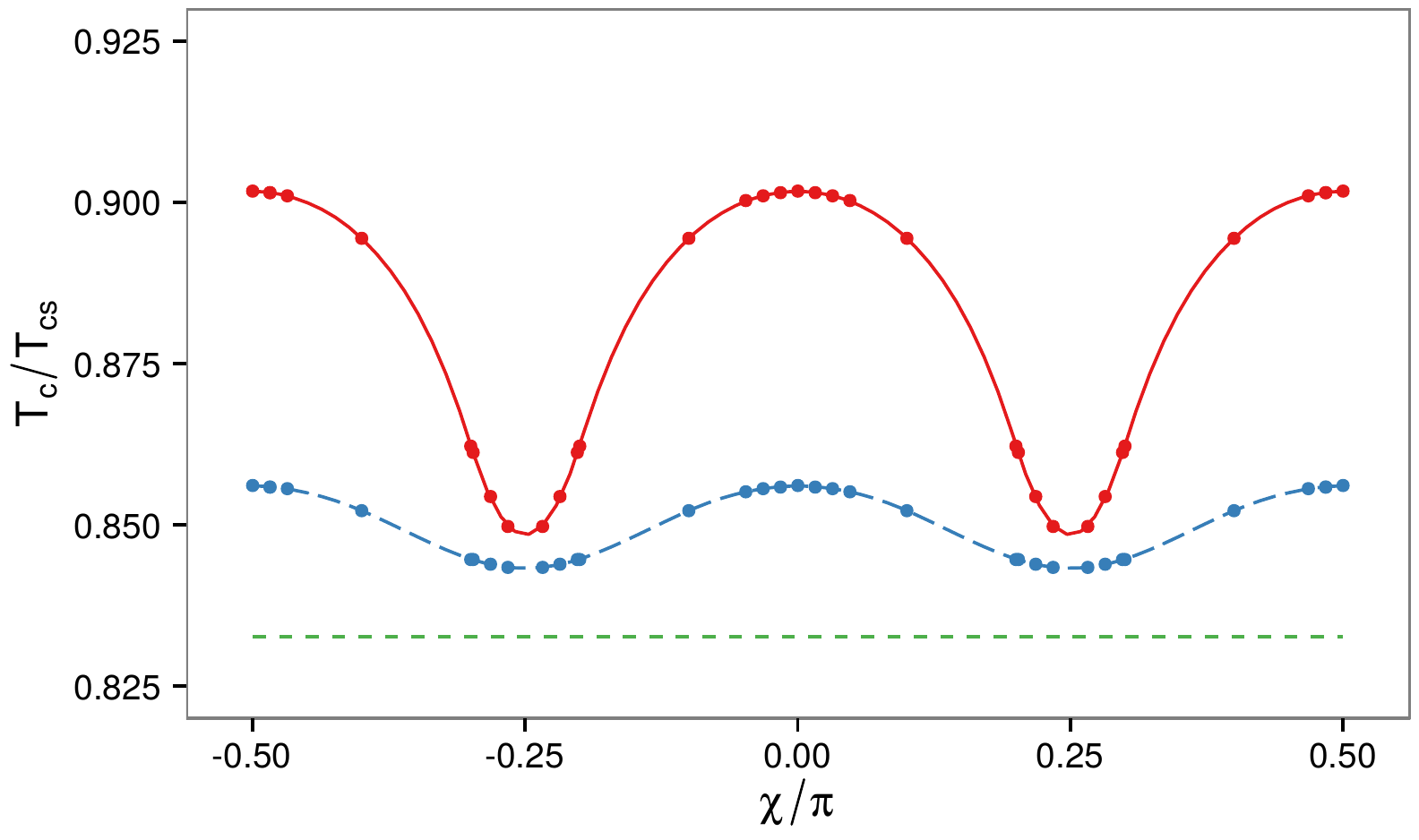}
	\caption{Plot of the critical temperature $T_c/T_{cs}$ as a function of the SO angle $\chi$, when $L_S/\xi_S = 1.00$, $L_F/\xi_S = 0.2$, and $\v h \parallel \uv x$. The critical temperature depends on the relative weight of the Rashba and Dresselhaus coefficients.}
	\label{fig:Tc-chi-x-long}
\end{figure}

The same situation for an exchange field along the $x$-axis is shown in \figrefs{Tc-chi-x-long}{Tc-chi-x-short}.
For this geometry, we observe a somewhat smaller critical temperature for all $a>0$ and all $\chi$ compared to \figrefs{Tc-chi-z-long}{Tc-chi-z-short}.
This can again be explained by considering the linearized Usadel equation in the ferromagnet, which suggests that the effective energy $E_x$ coupling to the $x$-component of the triplet vector should be
\begin{equation}
	E_x = \epsilon + 2iD_Fa^2 \; ,
\end{equation}
which has a smaller imaginary part than the corresponding equation for $E_z$.
Furthermore, note the drop in critical temperature as $\chi \rightarrow \pm \pi/4$.
Since the linearized equations contain a triplet mixing term proportional to $\sin 2\chi$, which is maximal precisely when $\chi = \pm\pi/4$, these are also the geometries for which we expect a maximal LRT generation.
Thus, this decrease in critical temperature near $\chi = \pm \pi/4$ can be explained by a net conversion of singlet components to LRTs in the system, which has an adverse effect on the singlet amplitude in the superconductor, and therefore the critical temperature.
\newpage
\begin{figure}[H]
	\centering
	\hspace{2.5em}
	\includegraphics{Tc_chi_labs.pdf}
	\hspace{-2em}
	\includegraphics[width=85mm]{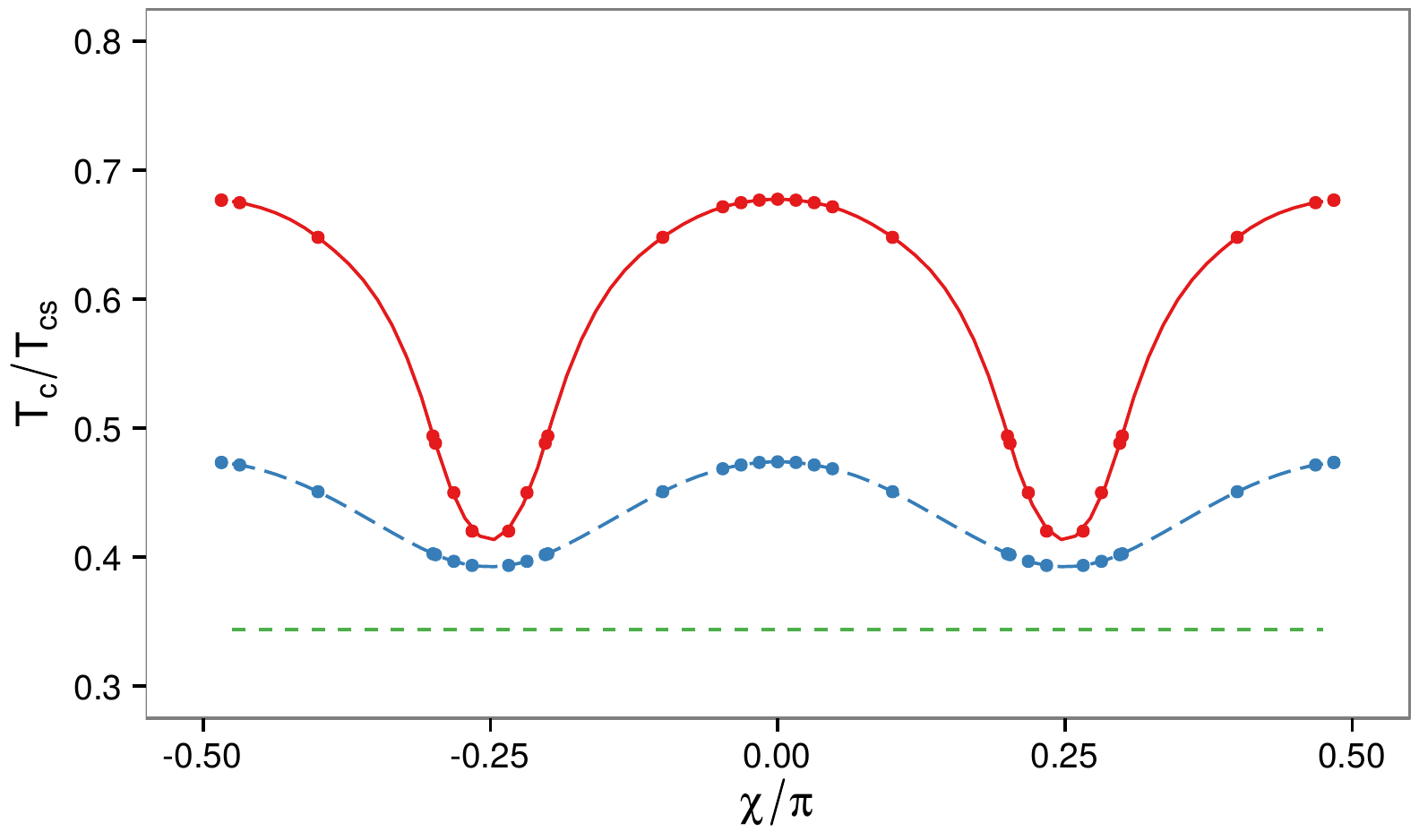}
	\caption{Plot of the critical temperature $T_c/T_{cs}$ as a function of the SO angle $\chi$, when $L_S/\xi_S = 0.55$, $L_F/\xi_S = 0.2$, and $\v h \parallel \uv x$. }
	\label{fig:Tc-chi-x-short}
\end{figure}

\begin{figure}[H]
	\centering
	\hspace{2.5em}
	\includegraphics{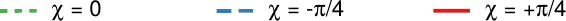}
	\hspace{-2em}
	\includegraphics[width=85mm]{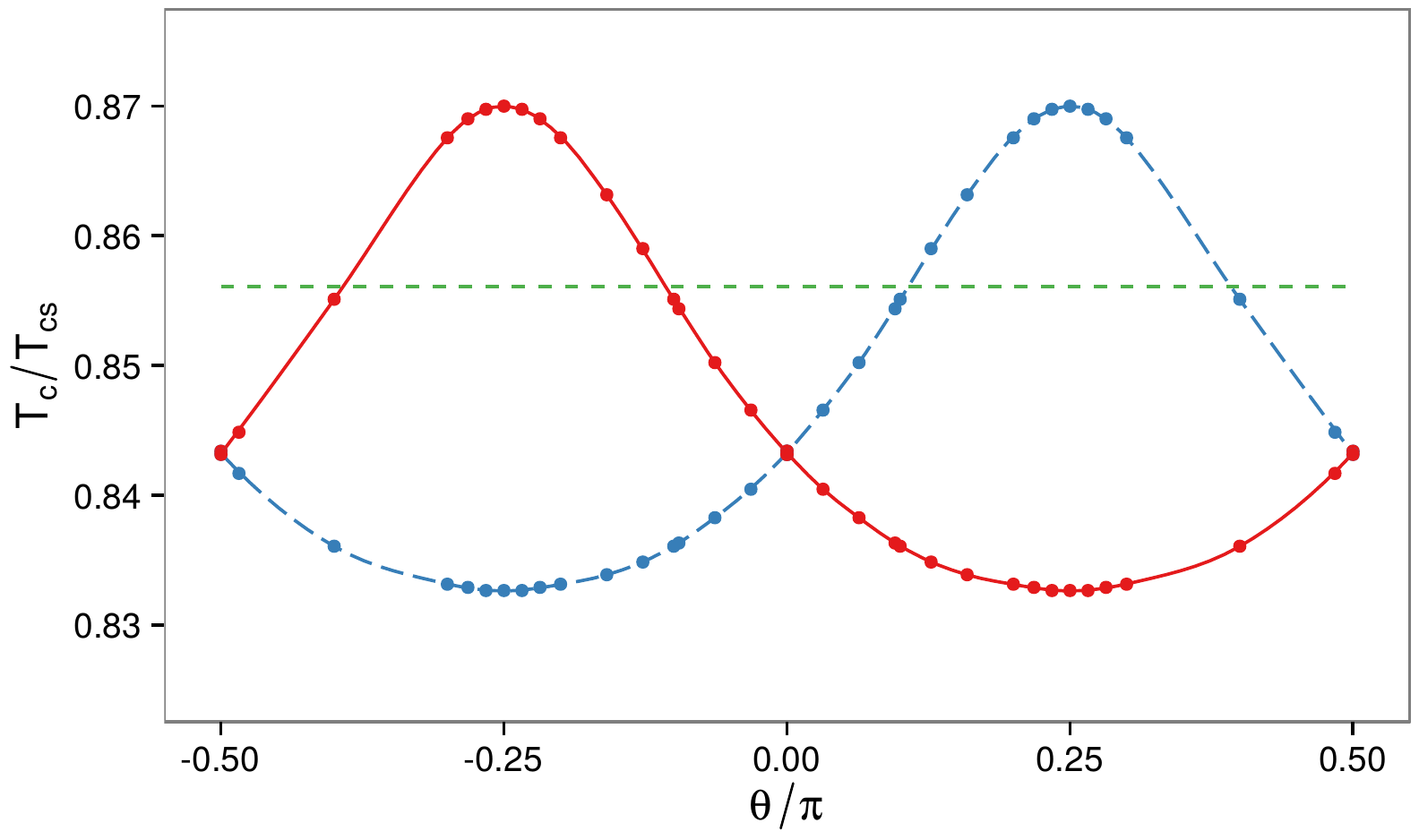}
	\caption{Plot of critical temperature $T_c/T_{cs}$ as a function of the exchange field angle $\theta$, when $L_S/\xi_S = 1.00$, $L_F/\xi_S = 0.2$, and $a\xi_S=2$. In contrast to ferromagnets without SO coupling, $T_c$ now depends strongly on the magnetization direction. This gives rise to a spin-valve like functionality with a single ferromagnet featuring SO coupling.}
	\label{fig:Tc-theta-long}
\end{figure}

\begin{figure}[H]
	\centering
	\hspace{2.5em}
	\includegraphics{theta_labs.pdf}
	\hspace{-2em}
	\includegraphics[width=85mm]{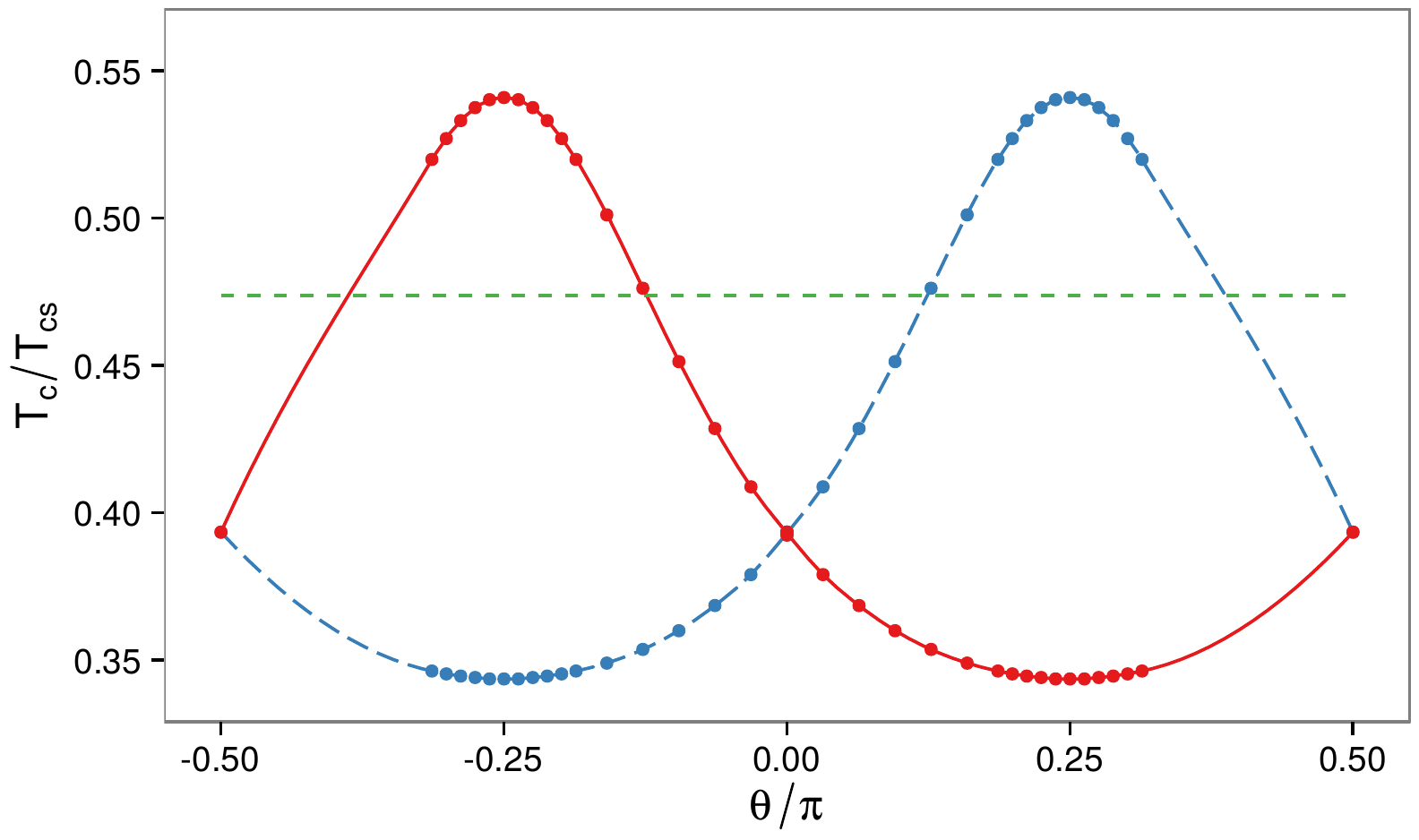}
	\caption{Plot of critical temperature $T_c/T_{cs}$ as a function of the exchange field angle $\theta$, when $L_S/\xi_S = 0.55$, $L_F/\xi_S = 0.2$, and $a\xi_S=2$.}
	\label{fig:Tc-theta-short}
\end{figure}

In \figrefs{Tc-theta-long}{Tc-theta-short} we present the results for a varying exchange field $\v h \sim \cos\theta\;\uv x + \sin\theta\;\uv y$ in the $xy$-plane.
In this case, we observe particularly interesting behaviour: the critical temperature has extrema at $\abs\chi=\abs\theta=\pi/4$, where the extremum is a maximum if $\theta$ and $\chi$ have the same sign, and a minimum if they have opposite signs.
Since $\theta = \pm\pi/4$ is precisely the geometries for which we do not expect any LRT generation, triplet mixing cannot be the source of this behaviour.
For the choice of physical parameters chosen in \figref{Tc-theta-short}, this effect results in a difference between the minimal and maximal critical temperature of nearly $60$\% as the magnetization direction is varied.
As shown in \figref{Tc-theta-long}, the effect persists qualitatively in larger structures as well, but is then weaker.\clearpage
\noindent
Instead, these observations may be explained using the theory developed in Section~\ref{Sec:Theory}.
When we have a general exchange field and SO field in the $xy$-plane, \eqnref{eq:E-SRT} reveals that the effective energy of the SRT component is
\begin{equation}
	E_\pll = \epsilon + 2iD_Fa^2 (1-\sin2\theta\;\sin2\chi) \; .
\end{equation}
Since the factor $(1-\sin2\theta\,\sin2\chi)$ vanishes for $\theta=\chi=\pm\pi/4$, we get $E_\pll = \epsilon$ for this case.
This geometry is also one where we do not expect any LRT generation, since the triplet mixing factor $\cos 2\theta\,\sin2\chi = 0$, so the conclusion is that the SO coupling has no effect on the behaviour of SRTs for these parameters---at least according to the linearized equations.
However, since $1-\sin2\theta\,\sin2\chi = 2$ for $\theta=-\chi=\pm\pi/4$, the situation is now dramatically different.
The SRT effective energy is now $E_\pll = \epsilon + 4iD_Fa^2$, with an imaginary contribution which again destabilizes the SRTs, and increases the critical temperature of the system. We emphasize that the variation of $T_c$ with the magnetization direction is present when $\chi \neq \pi/4$ as well, albeit with a magnitude of the variation that gradually decreases as one approaches pure Rashba or pure Dresselhais coupling.

\subsection{Triplet spin-valve effect with a single ferromagnet}
The results discussed in the previous section show that the critical temperature can be controlled via the magnetization direction of one single ferromagnetic layer. This is a new result originating from the presence of SO coupling. In conventional SF structures, $T_c$ is independent of the magnetization orientation of the F layer. By using a spin-valve setup such as FSF \cite{gu_prl_02, moraru_prl_06, zhu_prl_10, leksin_prl_12, banerjee_ncom_14}, it has been shown that the relative magnetization configuration between the ferromagnetic layers will tune the $T_c$ of the system. In contrast, in our case such a spin-valve effect can be obtained with a single ferromagnet (see Figs. \ref{fig:Tc-theta-long} and \ref{fig:Tc-theta-short}): by rotating the magnetization an angle $\pi/2$, $T_c$ goes from a maximum to a minimum. The fact that only a single ferromagnet is required to achieve this effect is of practical importance since it can be challenging to control the relative magnetization orientation in magnetic multilayered structures.

\section{Summary and discussion}\label{Sec:Disc}
It was pointed out in \refcite{BergeretTokatly2014} that for the case of transversal structures as depicted in Fig.~\ref{Fig:model}(b), pure Rashba or pure Dresselhaus coupling and arbitrary magnetisation direction are insufficient for long range triplets to exist. However, although these layered structures are more restrictive in their conditions for LRT generation than lateral junctions they are nevertheless one of the most relevant for current experimental setups \cite{KontosRyazanov,robinson,birge}, and herein we consider the corresponding experimentally accessible effects of SO coupling as a complement to the findings of \refcite{BergeretTokatly2014}.
We have provided a detailed exposition of the density of states and critical temperature for both the SF bilayer and SFS junction with SO coupling, highlighting in particular the signature of long range triplets.

We saw that the spectroscopic signature depends nonmonotonically on the angle of the magnetic exchange field, and that the LRT component can induce a strong peak in the density of states at zero energy for a range of magnetization directions.
In addition to the large enhancement at zero energy, we see that by carefully choosing the SO coupling and exchange field strengths in the Josephson junction it is again possible to control the qualitative features of the density of states by altering the phase difference between the two superconductors \eg with a loop geometry \cite{leSueuretal2008}.

The intrinsic SO coupling present in the structures considered herein derives from their lack of inversion symmetry due to the \eg junction interfaces, so-called interfacial asymmetry, and we restricted the form of this coupling to the experimentally common and, in some cases, tunable Rashba-Dresselhaus form.
A lack of inversion symmetry can also derive from intrinsic noncentrosymmmetry of a crystal.
This could in principle be utilised to provide a component of the SO-field in the junction direction, but to date we are not aware of such materials having been explored in experiments with SF hybrid materials. However, analytic and numerical data suggest that such materials could have significant importance for spintronic applications making use of a large triplet Cooper pair population\cite{JacobsenLinder2015}. 

It is also worth considering the possibility of separating the spin-orbit coupling and the ferromagnetic layer, which would arguably be easier to fabricate, and we are currently pursuing this line of investigation. In this case, we would expect similar conclusions regarding when the long-range triplets leave clear spectroscopic signatures and also regarding the spin-valve effect with a single ferromagnet, as found when the SO coupling and exchange field coexist in the same material. One way to practically achieve such a setup would be to deposit a very thin layer of a heavy normal metal such as Au or Pt between a superconductor and a conventional homogeneous ferromagnet. The combination of the large atomic number $Z$ and the broken structural inversion symmetry at the interface region would then provide the required SO coupling. With a very thin normal metal layer (of the order of a couple of nm), the proximity effect would be significantly stronger, and thus analysis of this regime is only possible with the full Usadel equations in the Riccati parameterisation developed herein.

The current analysis pertains to thin film ferromagnets.
Upon increasing the length of ferromagnetic film one will increase the relative proportions of long-range to short-range triplets in the middle of the ferromagnet. For strong ferromagnets where the exchange field is a significant fraction of the Fermi energy, the quasiclassical Usadel formalism may no longer describe the system behaviour appropriately, since it assumes that the impurity scattering rate is much larger than the other energy scales involved, and the Eilenberger equation should be used instead \cite{eilenberger}. 

In the previous section, we also observed that the presence of SO coupling will in many cases \emph{increase} the critical temperature of a hybrid structure.
This effect is explained through an increase in the effective energy coupled to the triplet component in the Usadel equation, which destabilizes the triplet pairs and closes that proximity channel.
However, for the special geometry $\theta = -\chi = \pm\pi/4$, the linearized equations suggest that the SRTs are unaffected by the presence of SO coupling, and this is consistent with the numerical results. 
We also note that for the geometries with a large LRT generation, such as $\theta = 0$ and $\chi = \pm\pi/4$, the LRT generation reduces the critical temperature again.
Thus, for the physical parameters considered herein, we see that there is a very slight increase in critical temperature for these geometries, but not as large as for the geometries without LRT generation.

One particularly striking result from the critical temperature calculations is that when the Rashba and Dresselhaus contribution to the SO coupling is of similar magnitude, one observes that the critical temperature can change by as much as 60\% upon changing $\theta=-\pi/4$ to $\theta=+\pi/4$, \ie by a 90$^\circ$ rotation of the magnetic field.
This implies that it is possible to create a novel kind of triplet spin valve using an SF bilayer, where the ferromagnet has a homogeneous exchange field and Rashba--Dresselhaus coupling. 
This is in contrast to previous suggestions for triplet spin valves, such as the one described by Fominov \etal, which have required trilayers with different homogeneous ferromagnets\cite{Fominov2010}.
The construction of such a device is likely to have possible applications in the emerging field of superconducting spintronics \cite{linder_nphys_15}.

\acknowledgments
The authors thank Angelo di Bernardo, Matthias Eschrig, Camilla Espedal, and Iryna Kulagina for useful discussions and gratefully acknowledge support from the `Outstanding Academic Fellows' programme at NTNU and COST Action MP-1201’ Novel Functionalities through Optimized
Confinement of Condensate and Fields’. J.L. was supported by the Research Council of Norway, Grant No. 205591 (FRINAT) and Grant No. 216700.

\appendix
\section{Riccati parametrization of the Usadel equation and Kupriyanov--Lukichev boundary conditions}\label{app:riccati}
The $4\times4$ components of the retarded Green's function~$\hat g$ are not entirely independent, but can be expressed as
\begin{align}
	\label{green-comp-form}
	\hat g(z,\epsilon) &= \begin{pmatrix} \phantom{-} g\,(z,\,+\epsilon) & \phantom{-} f\,(z,\,+\epsilon) \\ \!\!-\cc{f}\!(z,-\epsilon) & -\cc{g}\!(z,-\epsilon) \end{pmatrix} \, ,
\end{align} 
which suggests that the notation can be simplified by introducing the \emph{tilde conjugation}
\begin{equation}
	\tilde g(z,+\epsilon) \equiv \cc{g}\!(z,-\epsilon) \, .
\end{equation}
Moreover, the normalization condition $\hat g^2 = 1$ further constrains the possible form of $\hat g$ by relating the $g$ components to the $f$ components,
\begin{align}
	g g - f \tilde f &= 1 \, ,&
        g f - f \tilde g &= 0 \, .
\end{align}
Remarkably, if we pick a suitable parametrization of $\hat g$, which automatically satisfies the symmetry and normalization requirements above, then both the Usadel equation and the Kupriyanov--Lukichev boundary conditions can be reduced from $4\times4$ to $2\times2$ matrix equations.
In this paper, we employ the so-called \emph{Riccati parametrization} for this purpose, which is defined by
\begin{align}
  \label{riccati-par-def}
  \hat g = \begin{pmatrix} \rn & 0 \\ 0 & -\rnn \end{pmatrix} \begin{pmatrix} 1 + \rg\rgg & 2\rg \\ 2 \rgg & 1 + \rgg\rg \end{pmatrix} \, ,
\end{align}
where the normalization matrices are $\rn \equiv (1-\rg\rgg)^{-1}$ and $\rnn \equiv (1-\rgg\rg)^{-1}$.
Solving the Riccati parametrized equations for the function $\rg(z,\epsilon)$ in spin space is then sufficient to uniquely construct the whole Green's function $\hat g(z,\epsilon)$.
It is noteworthy that $\hat g \rightarrow 1$ when $\rg \rightarrow 0$, while the elements of $\hat g$ diverge to infinity when $\rg \rightarrow 1$; so we see that a finite range of variation in $\rg$ parametrizes an infinite range of variation in $\hat g$.

We begin by deriving some basic identities, starting with the inverses of the two matrix products $\rn\rg$ and $\rg\rnn$:
\begin{align}
	(\rn\rg)^{-1}  &= \rg^{-1}  \rn^{-1} = \rg^{-1} (1 - \rg\rgg) = \rg^{-1} - \rgg \; ; \\
	(\rg\rnn)^{-1} &= \rnn^{-1} \rg^{-1} = (1 - \rgg\rg) \rg^{-1} = \rg^{-1} - \rgg \; .
\end{align}
By comparison of the results above, we see that $\rn\rg = \rg\rnn$.
Similar calculations for other combinations of the Riccati matrices reveal that we can always move normalization matrices past gamma matrices if we also perform a tilde conjugation in the process:
\begin{equation}
	\begin{aligned}
		\rn\rg   &= \rg\rnn  \, , & 
		\rnn\rg  &= \rg\rn   \, , &
		\rn\rgg  &= \rgg\rnn \, , & 
		\rnn\rgg &= \rgg\rn  \, .
	\end{aligned}
\end{equation}
Since we intend to parametrize a differential equation, we should also try to relate the derivatives of the Riccati matrices.
This can be done by differentiating the definition of $\rn$ using the matrix version of the chain rule:
\begin{align}
	\partial_z \rn &= \partial_z (1-\rg\rgg)^{-1} \notag\\
                       &= -(1-\rg\rgg)^{-1} \left[ \partial_z(1-\rg\rgg) \right] (1-\rg\rgg)^{-1} \notag\\
                       &= (1-\rg\rgg)^{-1} \left[ \rgx\rgg + \rg\rggx \right] (1-\rg\rgg)^{-1} \notag\\
                       &= \rn \left[ \rgx\rgg + \rg\rggx \right] \rn \; .
\end{align}
Performing a tilde conjugation of the equation above, we get a similar result for $\partial_z\rnn$.
Thus, the derivatives of the normalization matrices satisfy the following identities:
\begin{align}
	\partial_z \rn  &= \rn  \left[ \rgx\rgg + \rg\rggx \right] \rn  \; ,\\
	\partial_z \rnn &= \rnn \left[ \rggx\rg + \rgg\rgx \right] \rnn \; .
\end{align}
In addition to the identities derived above, one should note that the definition of the normalization matrix $\rn = (1-\rg\rgg)^{-1}$ can be rewritten in many forms which may be of use when simplifying Riccati parametrized expressions; examples of this include $\rg\rgg = 1-\rn^{-1}$ and $1 = \rn-\rn\rg\rgg$.

Now that the basic identities are in place, it is time to parametrize the Usadel equation in the ferromagnet,
\begin{align}
	\label{eq:usadel-ferro}
	D_F\tilde{\nabla}(\hat{g}\tilde{\nabla} \hat{g})+i\left[\epsilon\hat{\rho}_3+\hat{M}, \hat{g}\right]=0 \, ,
\end{align}
where we for simplicity will let $D_F = 1$ in this appendix.
We begin by expanding the gauge covariant derivative $\cd(\hat g\cd\hat g)$, and then simplify the result using the normalization condition $\hat g^2 = 1$ and its derivative $\anticomm{\hat g}{\partial_z \hat g} = 0$, which yields the result
\begin{equation}
	\begin{aligned}
	\label{covder-exp-usadel}
		\cd\cdot(\hat g\cd\hat g) &= \partial_z(\hat g\partial_z \hat g) - i\partial_z(\hat g \hat A_z \hat g) \\
					  & - i\comm{\hat A_z}{\hat g\partial_z \hat g} -\comm{\hat A}{\hat g\hat A\hat g} \; .
	\end{aligned}
\end{equation}
We then write $\hat g$ in component form using \eqnref{green-comp-form}, and also write $\hat A$ in the same form using $\hat A = \mathrm{diag}(\v A, -\cc{\v A})$.
In the rest of this appendix, we will for simplicity assume that $\v A$ is real, so that $\hat A = \mathrm{diag}(\v A,-\v A)$; in practice, this implies that $\v A$ can only depend on the spin projections $\pauli x$ and $\pauli z$.
The derivation for the more general case of a complex $\hat A$ is almost identical.
The four terms in \eqnref{covder-exp-usadel} may then be written as follows:
\begin{align}
	& \partial_z(\hat g \partial_z \hat g)	\nonumber\\
	&= \begin{bmatrix} \phantom{-}\partial_z(       g \partial_z        g -        f \partial_z \tilde f) &
                                               \phantom{-}\partial_z(       g \partial_z        f -        f \partial_z \tilde g)	\phantom{-}\\
                                               \phantom{-}\partial_z(\tilde g \partial_z \tilde f - \tilde f \partial_z        g)	&
                                               \phantom{-}\partial_z(\tilde g \partial_z \tilde g - \tilde f \partial_z        f)	\phantom{-}
           \end{bmatrix} \; ; \\[1em]
	& \partial_z(\hat g \hat A \hat g)	\nonumber\\
	&= \begin{bmatrix}	\phantom{-}\partial_z(       g A        g +        f A \tilde f )			&
                                \phantom{-}\partial_z(       g A        f +        f A \tilde g )			\phantom{-}\\
                                	 - \partial_z(\tilde g A \tilde f + \tilde f A        g )			&
                                         - \partial_z(\tilde g A \tilde g + \tilde f A        f )			\phantom{-}
           \end{bmatrix} \; ; \\[1em]
	& \comm{\hat A}{\hat g \partial_z \hat g} \nonumber\\
	&= \begin{bmatrix} \phantom{-}\comm{    A}{       g \partial_z        g -        f \partial_z \tilde f\,}	&
                                                  		   \phantom{-}\anticomm{A}{       g \partial_z        f -        f \partial_z \tilde g}		\phantom{-}\\
                                                                            - \anticomm{A}{\tilde g \partial_z \tilde f - \tilde f \partial_z        g}		&
                                                                            - \comm{    A}{\tilde g \partial_z \tilde g - \tilde f \partial_z        f\,}	\phantom{-}
           \end{bmatrix} \; ; \\[1em]
	& \comm{\hat A}{\,\hat g \hat A \hat g} \nonumber\\
	&= \begin{bmatrix} \phantom{-}\comm{    A}{       g A        g +        f A \tilde f\,} 	&
       	  		   \phantom{-}\anticomm{A}{       g A        f +        f A \tilde g} 		\phantom{-}\\
                           \phantom{-}\anticomm{A}{\tilde g A \tilde f + \tilde f A        g} 		&
                           \phantom{-}\comm{    A}{\tilde g A \tilde g + \tilde f A        f\,} 	\phantom{-}
           \end{bmatrix} \; .
\end{align}
Substituting these results back into \eqnref{covder-exp-usadel}, we can find the upper blocks of the covariant derivative $\cd\cdot(\hat g\cd\hat g)$,
\begin{align}
	\label{dgdg-11}
		& [\cd\cdot(\hat g\cd \hat g)]^{(1,1)} 				\notag\\
		&= \partial_z(g\partial_z g-f\partial_z \tilde f)
        	 - i\partial_z(g A_z g + f A_z \tilde f) 			\notag\\
        	&- i\comm{A_z}{g\partial_z g - f\partial_z\tilde f}
        	 -  \comm{\v A}{g\v A g + f \v A \tilde f}  			\; ,  \\[1em]
	\label{dgdg-12}
		& [\cd\cdot(\hat g\cd \hat g)]^{(1,2)}				\notag\\
		&= \partial_z(g\partial_z f-f\partial_z \tilde g)
        	 - i\partial_z(g A_z f + f  A_z \tilde g) \notag		\notag\\
        	&- i\anticomm{A_z}{g\partial_z f - f\partial_z\tilde g}
        	 -  \anticomm{\v A}{g\v A f + f \v A \tilde g}			\; .
\end{align}
In this context, the notation $\hat M^{(n,m)}$ refers to the $n$'th row and $m$'th column in Nambu space.
Since the Green's function $\hat g$ and background field $\hat A$ also have a structure in spin space, the $(1,1)$ element in Nambu space is the upper-left $2\times2$ block of the matrix, and the $(1,2)$ element is the upper-right one.

There are two kinds of expressions that recur in the equations above, namely the components of $\hat g\partial_z\hat g$, and the components of $\hat g\hat A\hat g$.
After we substitute in the Riccati parametrization $g = 2\rn - 1$ and $f = 2\rn\rg$, these components take the form:
\begin{align}
	[\hat g \partial_z \hat g ]^{(1,1)}   &= g \partial_z g - f \partial_z \tilde f 		\notag\\
				     &= 2\rn \left[\rgx\rgg - \rg\rggx\right] \rn 			\; ;\\[0.5em]
	[\hat g \partial_z \hat g ]^{(1,2)}   &= g \partial_z f - f \partial_z \tilde g 		\notag\\
                                     &= 2\rn \left[\rgx - \rg\rggx\rg\right] \rnn  			\; ;\\[0.5em]
	[\hat g \hat A \hat g]^{(1,1)} &= g \v A g + f \v A \tilde f 					\notag\\
                                     &= 4\rn (\v A + \rg \v A \rgg)\rn - 2\anticomm{\v A}{\rn} + \v A 	\; ;\\[0.5em]
        [\hat g \hat A \hat g]^{(1,2)} &= g \v A f + f \v A \tilde g 					\notag\\
                                     &= 4\rn (\v A\rg + \rg \v A) \rnn - 2\anticomm{\v A}{\rn\rg}	\; .
\end{align}
If we explicitly calculate the commutators of $\hat A$ with the two matrices $\hat g\partial_z\hat g$ and $\hat g\hat A\hat g$, then we find:
\begin{align}
         \comm{\hat A}{\hat g\partial_z\hat g}^{(1,1)}
      &= \comm{\v A}{g\partial_z g - f \partial_z \tilde f}         			\notag\\
      &= 2\rn (1-\rg\rgg)\v A\rn \left[\rgx\rgg - \rg\rggx\right]  \rn 		\notag\\
      &- 2\rn \left[ \rgx\rgg - \rg\rggx \right] \rn\v A(1-\rg\rgg)\rn 		\; ;\\[0.5em]
         \comm{\hat A}{\hat g\partial_z\hat g}^{(1,2)}
      &= \anticomm{\v A}{g\partial_z f - f \partial_z \tilde g} 			\notag\\
      &= 2\rn (1-\rg\rgg)\v A\rn \left[\rgx - \rg\rggx\rg\right]    \rnn 		\notag\\
      &+ 2\rn \left[ \rgx - \rg\rggx\rg \right] \rnn\v A(1-\rgg\rg) \rnn 		\; ;\\[0.5em]
	 \comm{\hat A}{\hat g\hat A\hat g}^{(1,1)}
      &= \comm{\v A}{g\v Ag + f \v A \tilde f} 						\notag\\
      &= 4\v A\rn (\v A + \rg\v A\rgg) \rn						\notag\\
      &- 4\rn(\v A + \rg \v A \rgg) \rn \v A						\notag\\
      &- 2\comm{\v A^2}{\rn} 								\; ;\\[0.5em]
	\comm{\hat A}{\hat g\hat A\hat g}^{(1,2)}
     &= \anticomm{\v A}{g\v Af + f \v A \tilde g} 					\notag\\
     &= 4\v A\rn (\v A\rg + \rg\v A)\rnn 						\notag\\
     &+ 4\rn(\v A\rg + \rg\v A)\rnn\v A 						\notag\\
     &- 4\v A\rn\rg\v A - 2\anticomm{\v A^2}{\rn\rg}  					\; .
\end{align}
If we instead differentiate the aforementioned matrices with respect to $z$, we obtain:
\begin{align}
	[\partial_z(\hat g\partial_z \hat g)]^{(1,1)} &= \partial_z(g \partial_z g - f \partial_z \tilde f) 		\notag\\
                                                      &= 2\rn[ \rgxx + 2\rgx\rnn\rgg\rgx ] \rgg\rn 			\notag\\
                                                      &- 2\rn\rg[ \rggxx + 2\rggx \rn\rg\rggx ] \rn 			\; ;\\[1em]
	[\partial_z(\hat g\partial_z \hat g)]^{(1,2)} &= \partial_z(g \partial_z f - f \partial_z \tilde g) 		\notag\\
                                                      &= 2\rn[ \rgxx + 2\rgx\rnn\rgg\rgx ] \rnn 			\notag\\
                                                      &- 2\rn\rg[ \rggxx + 2\rggx\rn\rg\rggx ] \rg\rnn			\; ;\\[1em]
	 [\partial_z(\hat g\v A\hat g)]^{(1,1)}       &= \partial_z(g\v Ag + f \v A \tilde f) 				\notag\\
       						      &= 2\rn (1+\rg\rgg)\v A\rn[\rg\rggx + \rgx\rgg ] \rn		\notag\\
       						      &+ 2\rn[\rg\rggx + \rgx\rgg ] \rn\v A(1+\rg\rgg) \rn 		\notag\\
       						      &+ 4\rn \rg\v A\rnn[\rggx+\rgg\rgx\rgg] \rn			\notag\\
       						      &+ 4\rn[\rgx + \rg\rggx\rg]\rnn\v A\rgg \rn 			\; ;\\[1em]
	  [\partial_z(\hat g\v A\hat g)]^{(1,2)}      &= \partial_z(g\v Af + f \v A \tilde g) 				\notag\\
       						      &= 2\rn (1+\rg\rgg)\v A\rn[\rgx + \rg\rggx\rg ]   \rnn		\notag\\
       						      &+ 2\rn [\rgx + \rg\rggx\rg ] \rnn\v A(1+\rgg\rg) \rnn 	\notag\\
       						      &+ 4\rn \rg\v A\rnn[\rgg\rgx+\rggx\rg] \rnn			\notag\\
       						      &+ 4\rn [\rg\rggx + \rgx\rgg]\rn\v A\rgg \rnn 			\; .
\end{align}
Combining all of the equations above, we can express \eqnrefs{dgdg-11}{dgdg-12} using Riccati matrices.
In order to isolate the second-order derivative $\partial^2_z\gamma$ from these, the trick is to multiply \eqnref{dgdg-11} by $\rg$ from the right, and subsequently subtract the result from \eqnref{dgdg-12}:
\begin{align}
     \frac12 \rn^{-1} & \big\{ [\cd\cdot(\hat g\cd\hat g)]^{(1,2)} - [\cd\cdot(\hat g\cd\hat g)]^{(1,1)} \rg \big\} 	\notag\\
     &= \partial^2_z\gamma + 2\rgx\rnn\rgg\rgx 										\notag\\
     &-2i(A_z + \rg A_z \rgg)\rn\rgx 									     
      -2i\rgx\rnn( A_z + \rgg A_z\rg) 									     		\notag\\
     &-2(\v A\rg+\rg\v A)\rnn(\v A+\rgg\v A\rg) 								     
      -\v A^2\rg + \rg \v A^2 											     	\; .
\end{align}
If we finally rewrite $[\cd\cdot(\hat g\cd\hat g)]^{(1,1)}$ and $[\cd\cdot(\hat g\cd\hat g)]^{(1,2)}$ in the equation above by substituting in the Usadel equation~\eqref{eq:usadel-ferro}, then we obtain the following equation for the Riccati matrix~$\rg$:
\begin{align}
	\label{riccati-usadel}
	\partial^2_z\gamma =
	  &- 2i\epsilon\rg - i\v h\cdot(\v\sigma\rg - \rg\cc{\v\sigma}) -  2\rgx\rnn\rgg\rgx		\notag\\
          &+ 2i( A_z + \rg A_z\rgg)\rn\rgx 				
	   + 2i\rgx\rnn( A_z + \rgg A_z\rg) 								\notag\\
          &+ 2(\v A\rg + \rg \v A)\rnn(\v A + \rgg\v A\rg) 			
	   + \v A^2\rg - \rg\v A^2 									\; .
\end{align}
The corresponding equation for $\rgg$ can be found by tilde conjugation of the above.
After restoring the diffusion coefficient~$D_F$, and generalizing the derivation to a complex SO field $\v A$, the above result takes the form shown in \eqnref{Eqn:SOUsadel}.

After parametrizing the Usadel equation, the next step is to do the same to the Kupriyanov--Lukichev boundary conditions.
The gauge covariant version of \eqnref{Eqn:KL1} may be written
\begin{align}
	\label{eq:kl0}
	2L_n\zeta_n\hat{g}_n\cd\hat{g}_n=[\hat{g}_1,\hat{g}_2] \, ,
\end{align}
which upon expanding the covariant derivative $\hat g \cd \hat g$ becomes
\begin{align}
	\hat g_n \partial_z \hat g_n &=\frac{1}{2}\Omega_n\comm{\hat g_1}{\hat g_2} + i \hat g_n\comm{\hat A_z}{\hat g_n} \; ,
\end{align}
where we have introduced the notation $\Omega_n \equiv 1/L_n\zeta_n$ for the interface parameter.
We will now restrict our attention to the (1,1) and (1,2) components of the above,
\begin{align}
	g_n\partial_z g_n - f_n \partial_z \tilde f_n &= \frac12\Omega_n(g_1g_2 - g_2g_1 - f_1 \tilde f_2 + f_2 \tilde f_1) \notag\\
						      &\;+ ig_n\comm{ A_z}{g_n} + if_n\anticomm{ A_z}{\tilde f_n} 	    \; ,\\
	g_n\partial_z f_n - f_n \partial_z \tilde g_n &= \frac12\Omega_n(g_1f_2 - g_2f_1 - f_1 \tilde g_2 + f_2 \tilde g_1) \notag\\
						      &\;+ ig_n\anticomm{ A_z}{f_n} + if_n\comm{ A_z}{\tilde g_n}	     	    \; .
\end{align}
Substituting the Riccati parametrizations $g_n = 2\rrn n - 1$ and $f_n = 2\rrn n \rrg n$ in the above, we then obtain:
\begin{align}
	   \rrn n[\rrgx n \rrgg n - \rrg n \rrggx n] \rrn n	
	&=   \Omega_n \rrn1 (1-\rrg 1 \rrgg 2) \rrn2 				\notag\\
	&\,- \Omega_n \rrn2 (1-\rrg 2 \rrgg 1) \rrn1				\notag\\
	&\,- i \rrn n(1 - \rrg n \rrgg n) \v A \rrn n				\notag\\
	&\,- i \rrn n\v A(1 - \rrg n \rrgg n) \rrn n				\notag\\
	&\,+ 2i \rrn n (\v A + \rrg n \v A \rrgg n) \rrn n			\label{KL-11-comp} \; ,\\[1em]
	   \rrn n[\rrgx n - \rrg n \rrggx n \rrg n]\rrnn n			
	&=   \Omega_n \rrn1 (1-\rrg 1 \rrgg 2) \rrg2 \rrnn2 			\notag\\
	&\,- \Omega_n \rrn2 (1-\rrg 2 \rrgg 1) \rrg1	\rrnn1			\notag\\
	&\,+i\rrn n(1 + \rrg n \rrgg n)\v A \rrg n \rrnn n			\notag\\
	&\,+i\rrn n\rrg n \v A (1 + \rrgg n \rrg n)\rrnn n			\label{KL-12-comp} \; .
\end{align}
If we multiply \eqnref{KL-11-comp} by $\rrg n$ from the right, subtract this from \eqnref{KL-12-comp}, and divide by $\rrn n$ from the left, then we obtain the following boundary condition for  $\rrg n$:
\begin{align}
	\partial_z\rrg n &= \Omega_n (1 - \rrg 1 \rrgg 2) \rrn 2 (\rrg 2 - \rrg n)  	\notag\\
		          &\,+ \Omega_n (1 - \rrg 2 \rrgg 1) \rrn 1 (\rrg n - \rrg 1)	\notag\\
		          &\,+i\anticomm{ A_z}{\rrg n}				\; .
\end{align}
When we evaluate the above for $n=1$ and $n=2$, then it simplifies to the following:
\begin{align}
	\partial_z\rrg 1 &= \Omega_1 (1 - \rrg 1 \rrgg 2) \rrn 2 (\rrg 2 - \rrg 1) 
			  + i\anticomm{A_z}{\rrg 1} \label{riccati-kl-1}	    \; ,\\
	\partial_z\rrg 2 &= \Omega_2 (1 - \rrg 2 \rrgg 1) \rrn 1 (\rrg 2 - \rrg 1)
			  + i\anticomm{A_z}{\rrg 2} \label{riccati-kl-2}	    \; .
\end{align}
The boundary conditions for $\partial_z\rrgg 1$ and $\partial_z\rrgg 2$ are found by tilde conjugating the above.
If we generalize the derivation to a complex SO field $\v A$, and substitute back $\Omega_n \equiv 1/L_n\zeta_n$ in the result, then we arrive at \eqnref{Eqn:KLRic}.

\section{Derivation of the self-consistency equation for $\Delta$}\label{app:gap}
For completeness, we present here a detailed derivation of the self-consistency equation for the BCS order parameter \cite{bcs} in a quasiclassical framework. Similar derivations can also be found in Refs.~\onlinecite{kopnin, abri, Tinkham, serene, degennes}. In this paper, we follow the convention where the Keldysh component of the anomalous Green's function is defined as
\begin{equation}
	\label{def:keldysh-green}
	F^K_{\sigma\sigma'}(\v r,t;\,\v r',t') \equiv -i\expect{\comm{\p{\psi}_\sigma(\v r,t)}{\p{\psi}_{\sigma'}(\v r,t)}} \, ,
\end{equation}
where $\psi_\sigma(\v r,t)$ is the spin-dependent fermion annihilation operator, and the superconducting gap is defined as
\begin{equation}
	\label{def:gap}
	\Delta(\v r,t) \equiv \lambda\expect{ \p{\psi}_\up(\v r,t) \, \p{\psi}_\dn(\v r,t) } \, ,
\end{equation}
where $\lambda > 0$ is the electron--electron coupling constant in the BCS theory.
For the rest of this appendix, we will also assume that we work in an electromagnetic gauge where $\Delta$ is a purely real quantity.
Comparing \eqnrefs{def:keldysh-green}{def:gap}, and using the fermionic anticommutation relation
\begin{equation}
	\p{\psi}_\up(\v r,t) \, \p{\psi}_\dn(\v r,t) = -\p{\psi}_\dn(\v r,t) \, \p{\psi}_\up(\v r,t) \, ,
\end{equation}
we see that the superconducting gap $\Delta(\v r,t)$ can be expressed in terms of the Green's functions in two different ways,
\begin{align}
	\Delta(\v r,t) &= \phantom{-}\frac{i\lambda}{2} F^K_{\up\dn}(\v r,t;\,\v r,t) \, ,	\label{eq:gap-green-1} \\
	\Delta(\v r,t) &=          - \frac{i\lambda}{2} F^K_{\dn\up}(\v r,t;\,\v r,t) \, .	\label{eq:gap-green-2} 
\end{align}
We may then perform a quasiclassical approximation by first switching to Wigner mixed coordinates, then Fourier transforming the relative coordinates, then integrating out the energy dependence, and finally averaging the result over the Fermi surface to obtain the isotropic part.
The resulting equations for the superconducting gap are
\begin{align}
	\Delta(\v r,t) &= \phantom{-} \frac{1}{4} N_0\lambda \int \d\epsilon\, f^K_{\up\dn}(\v r,t,\epsilon) \, , \label{eq:gap-quasiclassical-1} \\
	\Delta(\v r,t) &=          -  \frac{1}{4} N_0\lambda \int \d\epsilon\, f^K_{\dn\up}(\v r,t,\epsilon) \, , \label{eq:gap-quasiclassical-2} 
\end{align}
where $f^K_{\sigma\sigma'}$ is the quasiclassical counterpart to $F^K_{\sigma\sigma'}$, $\epsilon$ is the quasiparticle energy, and $N_0$ is the density of states per spin at the Fermi level. 

In the equilibrium case, the Keldysh component $\hat g^K$ can be expressed in terms of the retarded and advanced components of the Green's function,
\begin{equation}
	\hat g^K = (\hat g^R - \hat g^A) \tanh(\epsilon/2T) \, ,
\end{equation}
and the advanced Green's function may again be expressed in terms of the retarded one,
\begin{equation}
	\hat g^A = -\rho_3 \hat g^{R\dagger} \rho_3 \, ,
\end{equation}
which implies that the Keldysh component can be expressed entirely in terms of the retarded component,
\begin{equation}
	\hat g^K = (\hat g^R - \rho_3 \hat g^{R\dagger} \rho_3) \tanh(\epsilon/2T) \, .
\end{equation}
If we extract the relevant anomalous components $f^K_{\up\dn}$ and $f^K_{\dn\up}$ from the above, we obtain the results
\begin{align}
	f^K_{\up\dn} &= [f^R_{\up\dn}(\v r,+\epsilon) + f^R_{\dn\up}(\v r,-\epsilon)] \tanh(\epsilon/2T) \, , \label{eq:keldysh-ra-1} \\
	f^K_{\dn\up} &= [f^R_{\dn\up}(\v r,+\epsilon) + f^R_{\up\dn}(\v r,-\epsilon)] \tanh(\epsilon/2T) \, . \label{eq:keldysh-ra-2}
\end{align}

We then switch to a singlet/triplet-decomposition of the retarded component $f^R$, where the singlet component is described by a scalar function $f_s$, and the triplet component by the so-called $d$-vector $(d_x,d_y,d_z)$. 
This parametrization is defined by the matrix equation
\begin{equation}
	f^R = (f_s + \v d \cdot \v \sigma) i\pauli y \, ,
\end{equation}
or in component form,
\begin{equation}
	\label{def:singlet-triplet-decomposition}
	\begin{pmatrix} f^R_{\up\up} &  f^R_{\up\dn} \vspace{0.2em}\\
                        f^R_{\dn\up} &  f^R_{\dn\dn} \end{pmatrix}
      = \begin{pmatrix} id_y - d_x &  d_z + f_s  \vspace{0.4em}\\
                         d_z - f_s & id_y + d_x  \end{pmatrix} \, .
\end{equation}
Parametrizing \eqnrefs{eq:keldysh-ra-1}{eq:keldysh-ra-2} according to \eqnref{def:singlet-triplet-decomposition}, we obtain
\begin{align}
	f^K_{\up\dn}(\v r,\epsilon) = [&d_z(\v r,+\epsilon) + f_s(\v r,+\epsilon)  			  	\nonumber\\ 
				     + &d_z(\v r,-\epsilon) - f_s(\v r,-\epsilon)] \tanh(\epsilon/2T) \, ,	\label{eq:kzs-1}\\
	f^K_{\up\dn}(\v r,\epsilon) = [&d_z(\v r,+\epsilon) - f_s(\v r,+\epsilon) 			  	\nonumber\\ 
                                     + &d_z(\v r,-\epsilon) + f_s(\v r,-\epsilon)] \tanh(\epsilon/2T) \, .	\label{eq:kzs-2}
\end{align}
The triplet component $d_z$ can clearly be eliminated from the above equations by subtracting \eqnref{eq:kzs-1} from \eqnref{eq:kzs-2},
\begin{equation}
	f^{K}_{\up\dn} - f^{K}_{\dn\up} = 2[f_s(\v r,\epsilon) - f_s(\v r,-\epsilon)] \tanh(\epsilon/2T) \, ,
\end{equation}
and a matching expression for the superconducting gap can be acquired by adding \eqnrefs{eq:gap-quasiclassical-1}{eq:gap-quasiclassical-2},
\begin{equation}
	2\Delta(\v r) = \frac14 N_0\lambda \int\d\epsilon\, [f^K_{\up\dn}(\v r,\epsilon)-f^K_{\up\dn}(\v r,\epsilon)] \tanh(\epsilon/2T) \, .
\end{equation}
By comparing the two results above, we finally arrive at an equation for the superconducting gap which only depends on the singlet component of the quasiclassical Green's function:
\begin{equation}
	\Delta(\v r) = \frac14 N_0\lambda \int\d\epsilon\, [f_s(\v r,\epsilon)-f_s(\v r,-\epsilon)] \tanh(\epsilon/2T) \, .
\end{equation}

If the integral above is performed for all real values of $\epsilon$, it turns out to be logarithmically divergent \eg for a bulk superconductor.
However, physically, the range of energies that should be integrated over is restricted by the energy spectra of the phonons that mediate the attractive electron--electron interactions in the superconductor.
This issue may therefore be resolved by introducing a Debye cutoff $\omega_c$, such that we only integrate over the region where $\abs{\epsilon} < \omega_c$.
Including the integration range, the gap equation is therefore
\begin{equation}
	\Delta(\v r) = \frac14 N_0\lambda \int\limits_{-\omega_c}^{\phantom{-}\omega_c} \d\epsilon\, [f_s(\v r,\epsilon)-f_s(\v r,-\epsilon)] \tanh(\epsilon/2T) \, .
\end{equation}
The equation above can however be simplified even further.
First of all, both $f_s(\epsilon)-f_s(-\epsilon)$ and $\tanh(\epsilon/2T)$ are clearly antisymmetric functions of $\epsilon$, which means that the product is a symmetric function, and so it is sufficient to perform an integral over positive values of $\epsilon$,
\begin{equation}
	\label{eq:gap-cutoff}
	\Delta(\v r) = \frac12 N_0\lambda \int\limits_{0}^{\phantom{-}\omega_c} \d\epsilon\, [f_s(\v r,\epsilon)-f_s(\v r,-\epsilon)] \tanh(\epsilon/2T) \, .
\end{equation}
However, because of the term $f_s(\v r,-\epsilon)$, we still need to know the Green's function for negative values of $\epsilon$ before we can calculate the gap.
On the other hand, the singlet component of the quasiclassical Green's functions also has a symmetry when the superconducting gauge is chosen as real
\begin{equation}
	f_s(\v r,\epsilon) = -f_s^*(\v r,-\epsilon) \, ,
\end{equation}
which implies that
\begin{equation}
	\label{eq:pm-symmetry}
	f_s(\v r,\epsilon) - f_s(\v r,-\epsilon) = 2\,\re{f_s(\v r,\epsilon)} \, .
\end{equation}
Substituting \eqnref{eq:pm-symmetry} into \eqnref{eq:gap-cutoff}, the gap equation takes a particularly simple form, which only depends on the real part of the singlet component $f_s(\v r,\epsilon)$ for positive energies $\epsilon$:
\begin{equation}
	\label{eq:gap-real}
	\Delta(\v r) = N_0\lambda \int\limits_{0}^{\phantom{-}\omega_c} \d\epsilon\; \re{f_s(\v r,\epsilon)} \tanh(\epsilon/2T) \, .
\end{equation}

Let us now consider the case of a BCS bulk superconductor, which has a singlet component given by the equation
\begin{equation}
	\label{eq:bulk-singlet}
	f_s(\epsilon) = \frac{\Delta}{\sqrt{\epsilon^2 - \Delta^2}} \, ,
\end{equation}
so that the gap equation may be written as
\begin{equation}
	\Delta = N_0 \lambda \int\limits_0^{\phantom{-}\omega_c} \d\epsilon\; \mathrm{Re}\left\{\frac{\Delta}{\sqrt{\epsilon^2-\Delta^2}}\right\} \tanh(\epsilon/2T) \, .
\end{equation}
The part in the curly braces is only real when $\abs{\epsilon}\geq\Delta$, which means that the equation can be simplified by changing the lower integration limit to $\Delta$.
After also dividing the equation by $\Delta N_0\lambda$, we then obtain the self-consistency equation
\begin{equation}
	 \frac{1}{N_0\lambda} = \int\limits_\Delta^{\phantom{-}\omega_c} \d\epsilon\; \frac{\tanh(\epsilon/2T)}{\sqrt{\epsilon^2-\Delta^2}} \, .
\end{equation}
For the zero-temperature case, where $T\rightarrow 0$ and $\Delta \rightarrow \Delta_0$, performing the above integral and reordering the result yields
\begin{equation}
	\label{eq:cutoff-result}
	\omega_c = \Delta_0 \cosh(1/N_0\lambda) \, .
\end{equation}
Using the above equation for $\omega_c$, and the well-known result
\begin{equation}
	\frac{\Delta_0}{T_c} = \frac{\pi}{e^\gamma} \, ,
\end{equation}
where $\gamma \approx 0.57722$ is the Euler--Mascheroni constant, we can finally rewrite \eqnref{eq:gap-real} as:
\begin{equation}
	\Delta(\v r) = N_0\lambda \int\limits_0^{\Delta_0 \cosh(1/N_0\lambda)\hspace{-5em}} \d\epsilon\; \re{f_s(\v r,\epsilon)} \tanh\left( \frac{\pi}{2e^\gamma} \frac{\epsilon/\Delta_0}{T/T_{c}} \right) \, .
\end{equation}
This version of the gap equation is particularly well-suited for numerical simulations.
One advantage is that we only need to know the Green's function for positive energies, which halves the number of energies that we need to solve the Usadel equation for.
The equation also takes a particularly simple form if we use energy units where $\Delta_0=1$ and temperature units where $T_c=1$, which is common practice in such simulations.

\end{document}